\documentclass[sigconf,authorversion,nonacm]{acmart}

\usepackage{tikz}
\usepackage{amsmath}
\usepackage{multirow}
\usepackage{balance}
\usepackage{lipsum}
\usepackage{xcolor}
\usepackage{listings}
\usepackage{enumitem}
\usepackage{booktabs}
\usepackage{comment}
\usepackage{makecell}
\usepackage{cleveref}
\usepackage{amsfonts}
\usepackage{pifont}
\usepackage{caption}
\usepackage{subcaption}
\usepackage[normalem]{ulem}

\definecolor{circlecolor}{HTML}{004d80}
\definecolor{redcirclecolor}{HTML}{a62b17}

\definecolor{n1_code}{HTML}{405063}
\definecolor{green}{HTML}{008800}
\definecolor{deepgreen}{rgb}{0,0.5,0}
\definecolor{props}{HTML}{db597c}
\definecolor{nicerblue}{HTML}{0066bb}
\definecolor{deepblue}{HTML}{0000dd}
\definecolor{velvetred}{HTML}{bb0066}

\definecolor{gray}{rgb}{0.5,0.5,0.5}
\definecolor{lg}{HTML}{DDDDDD}
\definecolor{CC}{HTML}{666666}
\definecolor{numbers}{HTML}{444444}

\definecolor{bgreen}{HTML}{20BC5F}

\usepackage[frozencache,cachedir=.]{minted}
\usepackage[breakable]{tcolorbox}
\definecolor{stateflow-pink}{HTML}{EFC9DB}
\definecolor{stateflow-orange}{HTML}{FFDD8C}
\definecolor{stateflow-blue}{HTML}{14A3D4}
\definecolor{stateflow-gray}{HTML}{F3F3F3}
\usepackage{booktabs} 
\usepackage[symbol]{footmisc}
\crefformat{section}{\S#2#1#3} 
\crefformat{subsection}{\S#2#1#3}
\crefformat{subsubsection}{\S#2#1#3}

\usepackage[ruled,vlined,linesnumbered]{algorithm2e}

\lstdefinestyle{pythonlang}{
  frame=single,
  language=Python,
  alsoletter={.},
  showstringspaces=false,
  emph={Hotel, Flight, \@reservation_operator.register, reserve, StatefulFunction, Operator, NotEnoughSpace},
  emphstyle={\color{velvetred}},
  columns=flexible,
  basicstyle={\fontfamily{lmtt}\scriptsize},
  identifierstyle={\color{black}},
  numbers=left,
  numberstyle=\tiny\color{deepblue},
  numbersep=5pt,
  keywordstyle=\color{blue},
  morekeywords={async, await, n_partitions, operator, function_name, key}, 
  commentstyle=\color{CC},
  stringstyle=\color{deepgreen},
  breaklines=false,
  breakatwhitespace=false,
  tabsize=1,
  xleftmargin=.0in,
  captionpos=b,
  keepspaces=true,
  escapechar=|
}

\newcommand{\entity}{\texttt{entity}}
\newcommand{\entities}{\texttt{entities}}
\newcommand{\functions}{\texttt{functions}}
\newcommand{\sstate}{\texttt{state}}
\newcommand{\key}{\texttt{key}}

\newcommand{\context}{\texttt{context}}

\newcommand*\circled[1]{\tikz[baseline=(char.base)]{
            \node[shape=circle,line width=0mm,inner sep=1pt,fill = circlecolor,text=white] (char) {\textsf{{#1}}};}}

\newcommand*\circler[1]{\tikz[baseline=(char.base)]{
            \node[shape=circle,line width=0mm,inner sep=1pt,fill = redcirclecolor,text=white] (char) {\textsf{{#1}}};}}

\newcommand{\para}[1]{\vspace{1.5mm}\noindent\textbf{#1.}}
\newcommand{\parait}[1]{\vspace{0mm}\noindent\textit{\underline{#1.}}}

\newcommand{\sysname}{Styx}

\usepackage[colorinlistoftodos]{todonotes}

\newcommand{\rev}[1]{{\color{black} #1}}

\begin{document}

\title{Styx: Transactional Stateful Functions on Streaming Dataflows}

\author{Kyriakos Psarakis}
\affiliation{%
  \institution{Delft University of Technology}
  \country{}
}
\email{k.psarakis@tudelft.nl}

\author{George Christodoulou}
\affiliation{%
  \institution{Delft University of Technology}
  \country{}
}
\email{g.c.christodoulou@tudelft.nl}

\author{George Siachamis}
\affiliation{%
  \institution{Inria \& Institut Polytechnique de Paris}
  \country{}
}
\email{georgios.siachamis@inria.fr}
\authornote{Work done while at Delft University of Technology}

\author{Marios Fragkoulis}
\affiliation{%
  \institution{Delft University of Technology}
  \country{}
}
\email{m.fragkoulis@tudelft.nl}

\author{Asterios Katsifodimos}
\affiliation{%
  \institution{Delft University of Technology}
  \country{}
}
\email{a.katsifodimos@tudelft.nl}

\renewcommand{\shortauthors}{Psarakis et al.}


\begin{abstract}
Developing stateful cloud applications, such as low-latency workflows and microservices with strict consistency requirements, remains arduous for programmers. The Stateful Functions-as-a-Service (SFaaS) paradigm aims to serve these use cases. However, existing approaches provide weak transactional guarantees or perform expensive external state accesses requiring inefficient transactional protocols that increase execution latency. 

In this paper, we present \sysname{}, a novel dataflow-based SFaaS runtime that executes serializable transactions consisting of stateful functions that form arbitrary call-graphs with exactly-once guarantees. \sysname{} extends a deterministic transactional protocol by contributing: i) a function acknowledgment scheme to determine transaction boundaries required in SFaaS workloads, ii) a function-execution caching mechanism, and iii) an early-commit reply mechanism that substantially reduces transaction execution latency. Experiments with the YCSB, TPC-C, and Deathstar benchmarks show that \sysname{} outperforms state-of-the-art approaches by achieving at least one order of magnitude higher throughput while exhibiting near-linear scalability and low latency.
\end{abstract}

\maketitle

\section{Introduction}

Despite the commercial offerings of the Functions-as-a-Service (FaaS) cloud service model, its suitability for low-latency stateful applications with strict consistency requirements, such as payment processing, reservation systems, inventory keeping, and low-latency business workflows, is quite limited. The reason behind this unsuitability is that current FaaS solutions are stateless, relying on external, fault-tolerant data stores (blob stores or databases) for state management. In addition, while multiple frameworks can perform workflow execution (e.g., AWS Step Functions \cite{stepfunctions}, Azure Logic Apps \cite{logicapps}), they do not provide primitives for \textit{transactional} execution of such applications. 
As a result, distributed applications (e.g., microservice architectures) suffer from serious consistency issues when the responsibility of transaction execution is left to developers \cite{blanastransactions,microservices-survey}.

In line with recent research \cite{cloudburst, boki, beldi, tstatefun, nightcore, spenger2022portals}, we agree that for FaaS offerings to become mainstream, they should include state management support for stateful functions according to the Stateful Functions-as-a-Service (SFaaS) paradigm. In addition, we argue that a suitable runtime for executing workflows of stateful functions should also provide $i)$ end-to-end serializable transactional guarantees across multiple functions, $ii)$ low-latency and high-throughput execution, and $iii)$ a high-level programming model, devoid of low-level primitives for locking and transaction coordination. To the best of our knowledge, no existing approach addresses all these requirements together.

\begin{figure}[t]
    \centering
    \includegraphics[width=\columnwidth]{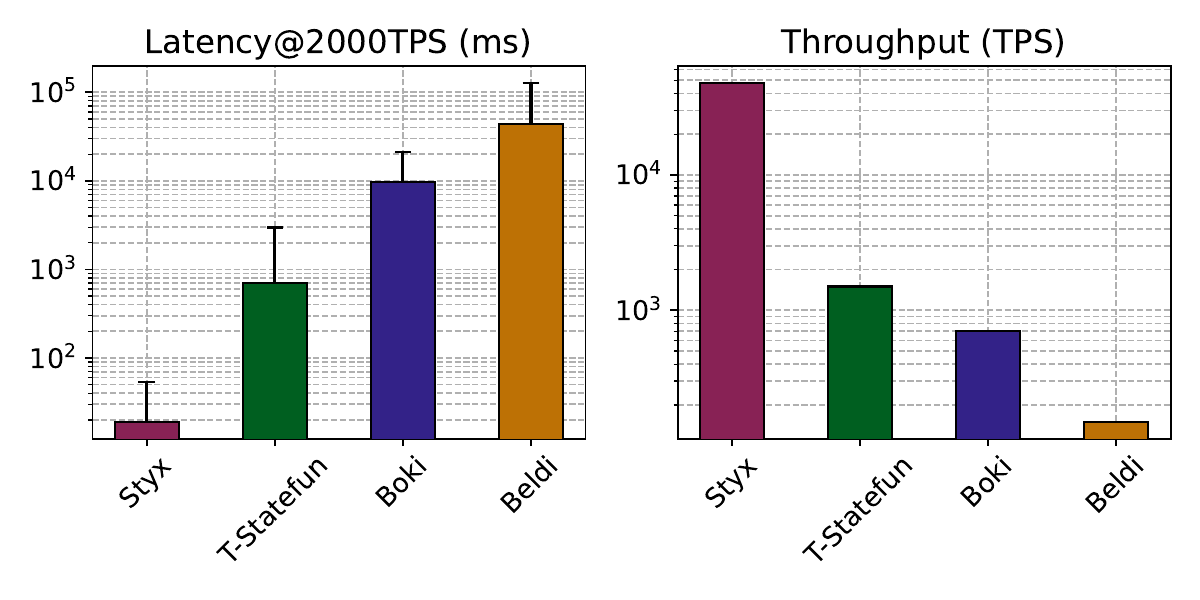}
    \vspace{-9mm}
    \caption{\sysname{} outperforms the SotA by at least one order of magnitude in transactional workloads (\Cref{sec:exp}). The figure shows median (bar)/99p (whisker) latency and throughput. For the latency plot, the input throughput is 2000 transactions per second (TPS), and for the throughput plot, we report the throughput that the systems achieve at subsecond latency.}
    \label{fig:summary_comparison}
    \vspace{-5mm}
\end{figure}



The state-of-the-art transactional SFaaS with serializable guarantees, Boki~\cite{boki}, Beldi~\cite{beldi}, and T-Statefun~\cite{tstatefun} do support transactional end-to-end workflows but induce high commit latency and low throughput. The main reason behind their inefficiency is the separation of state storage and function logic, as well as the use of locking and Two-Phase Commit (2PC)~\cite{2pc} to coordinate and ensure the atomicity of cross-function transactions.






This paper proposes \sysname{}, a novel dataflow-based runtime for SFaaS. \sysname{} ensures that each transaction's state mutations will be reflected once in the system's state, even under failures, retries, or other potential disruptions (known as exactly-once processing).
Additionally, \sysname{} can execute arbitrary function orchestrations with end-to-end serializability guarantees, leveraging concepts from deterministic databases to avoid costly 2PCs. 

Our work stems from two important observations. First, modern streaming dataflow systems such as Apache Flink \cite{flink} guarantee exactly-once processing \cite{flink,CarboneEF17,silvestre2021clonos} by transparently handling failures. A limitation of those streaming systems is that they cannot execute general cloud applications such as microservices or guarantee transactional SFaaS orchestrations. Second, deterministic database protocols \cite{thomson2012calvin,aria} that can avoid expensive 2PC invocations have not been designed for complex function orchestrations and arbitrary call-graphs. For the needs of transactional SFaaS, \sysname{} leverages a deterministic transactional protocol, enabling early commit replies to clients (i.e., before a snapshot is committed to persistent storage).

Our work is in line with recent proposals in the area, such as DBOS~\cite{dbos}, Hydro~\cite{newdirectionscloud}, and SSMSs~\cite{li2024serverless}. Contrary to these systems, our work adopts the streaming dataflow execution model and guarantees serializability \textit{across} functions. As shown in \Cref{fig:summary_comparison}, \sysname{} achieves one order of magnitude lower median latency, two orders of magnitude lower 99p latency at 2000 transactions/sec, and one order of magnitude higher throughput compared to state of the art (SotA) serializable SFaaS systems \cite{boki, beldi, tstatefun}.

\vspace{2mm}

\noindent In short, this paper makes the following contributions:

\rev{\noindent \textbf{--} \sysname{} combines deterministic transactions with dataflows and overcomes the challenges that arise from this design choice (\Cref{sec:background}).

\noindent \textbf{--} \sysname{} enables high-level SFaaS programming models that abstract away transaction and failure management code (\Cref{sec:programming-model}). \sysname{} does so, by guaranteeing exactly-once processing (\Cref{sec:dataflow-system}) and transactional serializability across arbitrary function calls (\Cref{sec:seq_f_ex} and \Cref{sec:cmt_t}).

\noindent \textbf{--} \sysname{} extends the concept of deterministic databases to support arbitrary workflows of stateful functions, contributing a novel acknowledgment scheme (\Cref{sec:ack_shares}) to track function completion efficiently, as well as a function-execution caching mechanism (\Cref{sec:caching}) to speed up function re-executions.

\noindent \textbf{--} \sysname{}'s deterministic execution enables early commit replies: transactions can be reported as committed, even before a snapshot of executed transactions is committed to durable storage (\Cref{sub:early}).

\noindent \textbf{--} \sysname{} outperforms the state-of-the-art \cite{beldi,boki,tstatefun} by at least one order of magnitude higher throughput in all tested workloads while achieving lower latency and near-linear scalability (\Cref{sec:exp}).}


\vspace{2mm}
\noindent \sysname{} is available at: \url{https://github.com/delftdata/styx}
\vspace{-1mm}
\section{Motivation} \label{sec:background}
In this section, we analyze the specifics of streaming dataflow systems design and argue that they can be extended to encapsulate the primitives required for consistently and efficiently executing workflows of stateful functions.
\rev{Our work is based on a key observation: the architecture of high-performance  cloud services closely resembles a parallel dataflow graph, where the state is partitioned and co-located with the application logic~\cite{styxcidr}.
Additionally, as we detail in \Cref{sec:determinism-transactions}, there is a synergy between deterministic transactions and dataflow systems. Such a combination can offer state consistency and ease of programming as monolithic solutions did in the past, while improving scalability and eliminating developer involvement. Finally, we show how deterministic transactions can be extended for SFaaS, where transaction boundaries are unknown, unlike online transaction processing (OLTP).}

\vspace{-1mm}
\subsection{Dataflows for Stateful Functions}

Stateful dataflows is the execution model implemented by virtually all modern stream processors \cite{flink,murray2013naiad,NoghabiPP17}. Besides being a great fit for parallel, data-intensive computations, stateful dataflows are the primary abstraction supporting workflow managers such as Apache Airflow \cite{airflow}, AWS Step Functions \cite{stepfunctions}, and Azure's Durable Functions \cite{burckhardt2021durable}. In the following, we present the primary motivation behind using stateful dataflows to build a suitable runtime for orchestrating general-purpose cloud applications. 

\para{Exactly-once Processing} Message-delivery guarantees are fundamentally hard to deal with in the general case, with the root of the problem being the well-known Byzantine Generals problem \cite{lamport1982byzantine}. However, in the closed world of dataflow systems, exactly-once processing is possible \cite{flink,CarboneEF17,silvestre2021clonos}. As a matter of fact, the APIs of popular streaming dataflow systems, such as Apache Flink, require no error management code (e.g., message retries or duplicate elimination with idempotency IDs).

\para{Co-Location of State and Function} The primary reason streaming dataflow systems can sustain millions of events per second \cite{flink,jet} is that their state is partitioned across operators that operate on local state. While the structure of current Cloud offerings favors the disaggregation of storage and computation, we argue that co-locating state and computation is the primary vehicle for high performance and can also be adopted by modern SFaaS runtimes, as opposed to using external databases for state storage.

\para{Coarse-Grained Fault Tolerance} To ensure atomicity at the level of workflow execution, existing SFaaS systems perform fine-grained fault tolerance \cite{beldi,boki}; each function execution is logged and persisted in a shared log before the next function is called. This requires a round-trip to the logging mechanism for each function call, which adds significant latency to function execution. Instead of logging each function execution, streaming dataflow systems \cite{checkmate,CarboneEF17,chandy1985distributed} opt for a coarse-grained fault tolerance mechanism based on asynchronous snapshots, reducing this overhead.

\vspace{-2mm}
\subsection{Determinism \& Transactions}
\label{sec:determinism-transactions}

Given a set of database partitions and a set of transactions, a deterministic database \cite{Abadi2018Deterministic, Thomson2010Calvin} will end up in the same final state despite node failures and possible concurrency issues. Traditional database systems offer \textit{serializable} guarantees, allowing multiple transactions to execute concurrently, ensuring that the database state will be equivalent to the state of one serial transaction execution. Deterministic databases guarantee not only serializability but also that a given set of transactions will have exactly the same effect on the database state despite transaction re-execution. 
This guarantee has important implications \cite{Abadi2018Deterministic} that have not been leveraged by SFaaS systems thus far. 

\para{Deterministic Transactions on Streaming Dataflows} Unlike 2PC, which requires rollbacks in case of failures, deterministic database protocols \cite{aria,thomson2012calvin} are "forward-only": once the locking order \cite{thomson2012calvin} or read/write set \cite{aria} of a batch of transactions has been determined, the transactions are going to be executed and reflected on the database state, without the need to rollback changes. This notion is in line with how dataflow systems operate: events flow through the dataflow graph, from sources to sinks, without stalls for coordination. This match between deterministic databases and the dataflow execution model is the primary motivation behind \sysname{}'s design choice to implement a deterministic transaction protocol on top of a dataflow system.

\subsection{Challenges} 
Despite their success and widespread applicability, dataflow systems need to undergo multiple changes before they can be used for transactional stateful functions. In the following, we list challenges and open problems tackled in this work.

\para{Programming Models} Dataflow systems at the moment are only programmable through functional programming-style dataflow APIs: a given cloud application has to be rewritten by programmers to match the event-driven dataflow paradigm. Although it is possible to rewrite many applications in this paradigm, it takes a considerable amount of programmer training and effort. We argue that dataflow systems would benefit from object-oriented or actor-like programming abstractions in order to be adopted for general cloud applications, such as microservices.

\para{Support for Transactions} Transactions in the context of streaming dataflow systems typically refer to processing a set of input elements and their state updates with ACID guarantees \cite{zhang2024survey}. Despite progress, critical challenges remain open, such as the performance overhead incurred by multi-partition transactions, as well as the need to block flows of data for locking and message re-ordering. In this work, we argue that in order to implement transactions in a streaming dataflow system, we need to "keep the data moving" \cite{stonebraker20058} by avoiding disruptions in the natural flow of data while tightly integrating transaction processing into the system's state management and fault tolerance protocols.


\para{Deterministic OLTP and SFaaS} OLTP databases that use deterministic protocols like Calvin~\cite{thomson2012calvin, zhou2022lotus, aria} either require each transaction's read/write set a priori or are extended to discover the read-write sets of a transaction by first executing it. Additionally, in both scenarios, deterministic protocols assume that a transaction is executed as a single-threaded function that can perform remote reads and writes from other partitions.
In the case of SFaaS, arbitrary function calls enable programmers to take advantage of both the separation of concerns principle, which is widely applied in microservice architectures \cite{microservices-survey}, as well as code modularity. Although deterministic database systems have been proven to perform exceptionally well~\cite{Abadi2018Deterministic}, designing and implementing a deterministic transactional protocol for arbitrary workflows of stateful functions is non-trivial. Specifically, arbitrary function calls create complex call-graphs that need to be tracked in order to establish a transaction's boundaries before committing.

\para{Dataflows for Arbitrary-Workflow Execution} The prime use case for dataflow systems nowadays is streaming analytics. However, general-purpose cloud applications have different workload requirements. Functions calling other functions and receiving responses introduce cycles in the dataflow graph. Such cycles can cause deadlocks and need to be dealt with \cite{faucet}.

\vspace{2mm}

\noindent In this work, we tackle these challenges and propose a dataflow system tailored to the needs of stateful functions with built-in support for deterministic transactions and a high-level programming model.


\section{Programming Model} \label{sec:programming-model}

The programming model of \sysname{} is based on Python and comprises operators that encapsulate partitioned mutable state and functions that operate on that. An example of the programming model of \sysname{} is depicted in \Cref{fig:programming-model}.

\vspace{-2mm}

\subsection{Programming Model Notions}

\para{Stateful Entities} Similar to objects in object-oriented programming, \entities{} in \sysname{} are responsible for maintaining and mutating their own \sstate{}. Moreover, when a given entity needs to update the state of another entity, it can do so via a function call. Each entity bears a unique and immutable \key{}, similar to Actor references in Akka~\cite{Akka}, with the difference that entity keys are application-dependent and contain no information related to their physical location. The dataflow runtime engine (\Cref{sec:dataflow-system}) uses that key to route function calls to the right operator that accommodates that specific \entity{}. 

\para{Functions} \functions{} can mutate the state of an entity. By convention, the \context{} is the first parameter of each function call. Functions are allowed to call other functions directly, and \sysname{} supports both synchronous and asynchronous function calls. For instance, in lines 9-11 of \Cref{fig:programming-model}, the instantiated reservation entity will call asynchronously the function \texttt{'reserve\_hotel'} of an entity with key \texttt{'hotel\_id'} attached to the Hotel operator. Similarly, one can make a synchronous call that blocks waiting for results. In this case, \sysname{} will block execution until the call returns. Depending on the use case, a mix of synchronous and asynchronous calls can be used. Asynchronous function calls, however, allow for further optimizations that \sysname{} applies whenever possible, as we describe in \Cref{sec:seq_f_ex} and \Cref{sec:cmt_t}.

\begin{figure}[t]
\begin{lstlisting}[style=pythonlang]
from styx import Operator
from deathstar.operators import Hotel, Flight

reservation_operator = Operator('reservation', n_partitions=4)

@reservation_operator.register
async def make_reservation(context, flight_id, htl_id, usr_id):

    context.call_async(operator=Hotel,
                       function_name='reserve_hotel',
                       key=htl_id)
    context.call_async(operator=Flight,
                       function_name='reserve_flight',
                       key=flight_id)

    reservation = {"fid":flight_id, "hid":htl_id, "uid":usr_id}
    await context.state.put(reservation)|\label{code:context}|
    
    return "Reservation Successful"
\end{lstlisting}
\vspace{-3mm}
\caption{Deathstar's\cite{deathstar} Hotel/Flight reservation in \sysname{}. From lines 9-14, the $reserve\_hotel$ and $reserve\_flight$ functions are invoked asynchronously. Finally, in lines 16-17, the reservation information is stored. In \sysname{}, the transactional and fault tolerance logic are handled internally.}
\label{fig:programming-model}
\vspace{-3mm}
\end{figure}

\para{Operators} Each \entity{} directly maps to a dataflow operator (also called a vertex) in the dataflow graph. When an \textit{event} enters the dataflow graph, it reaches the operator holding the \textit{function code} of the given entity as well as the \textit{state} of that entity. In short, a dataflow operator can execute all functions of a given entity and store the state of that entity. Since operators can be partitioned across multiple cluster nodes, each partition stores a set of stateful entities indexed by their unique \key{}. When an entity's function is invoked (via an incoming event), the entity's state is retrieved from the local operator state. Then, the function is executed using the arguments found in the incoming event that triggered the call. 

\para{State \& Namespacing} As mentioned before, each entity has access only to its own state. In \sysname{}, the state is \emph{namespaced} with respect to the entity it belongs to. For instance, a given key "\texttt{hotel53}" within the operator \texttt{Hotel} is represented as: \texttt{entities://Hotel/hotel53}. This way, a reference to a given key of a state object is unique and can be determined at runtime when operators are partitioned across workers. Programmers can store or retrieve \sstate{} through the \texttt{context} object by invoking \texttt{context.put()} or \texttt{get()} (e.g., in \Cref{code:context} of \Cref{fig:programming-model}). \sysname{}'s \texttt{context} is similar to the context object used in other systems such as Flink Statefun, AWS Lambda, and Azure Durable Functions.

\para{Transactions} A transaction in \sysname{} begins with a client request. The functions that are part of the transaction form a workflow that executes with serializable guarantees. \sysname{}'s programming model allows transaction aborts by raising an uncaught exception. In the example of \Cref{fig:programming-model}, if a hotel entity does not have enough availability when calling the \textit{'reserve\_hotel'} function, the \textit{'make\_reservation'} transaction should be aborted, alongside potential state mutations that the \textit{'reserve\_flight'} has made to a flight entity. In that case, the programmer has to raise an exception as follows:

\begin{lstlisting}[style=pythonlang]
...
# Check if there are enough rooms available in the hotel
if available_rooms <= 0: 
    raise NotEnoughSpace(f'No rooms in hotel: {context.key}')
...
\end{lstlisting}
The exception is caught by \sysname{}, which automatically triggers the abort/rollback sequence of the transaction where the exception occurred and sends the user-defined exception message as a reply.

\para{Exactly-once Function Calling}
\sysname{} offers \emph{exactly-once processing} guarantees: it reflects the state changes of a function call execution exactly-once. Thus, programmers do not need to ``pollute'' their application logic with consistency checks, state rollbacks, timeouts, retries, and idempotency \cite{microservices-survey, microservices-drawbacks}. We detail this capability in \Cref{sec:fault-tolerance}.

\begin{figure}[t]
    \centering
    \includegraphics[width=0.85\columnwidth]{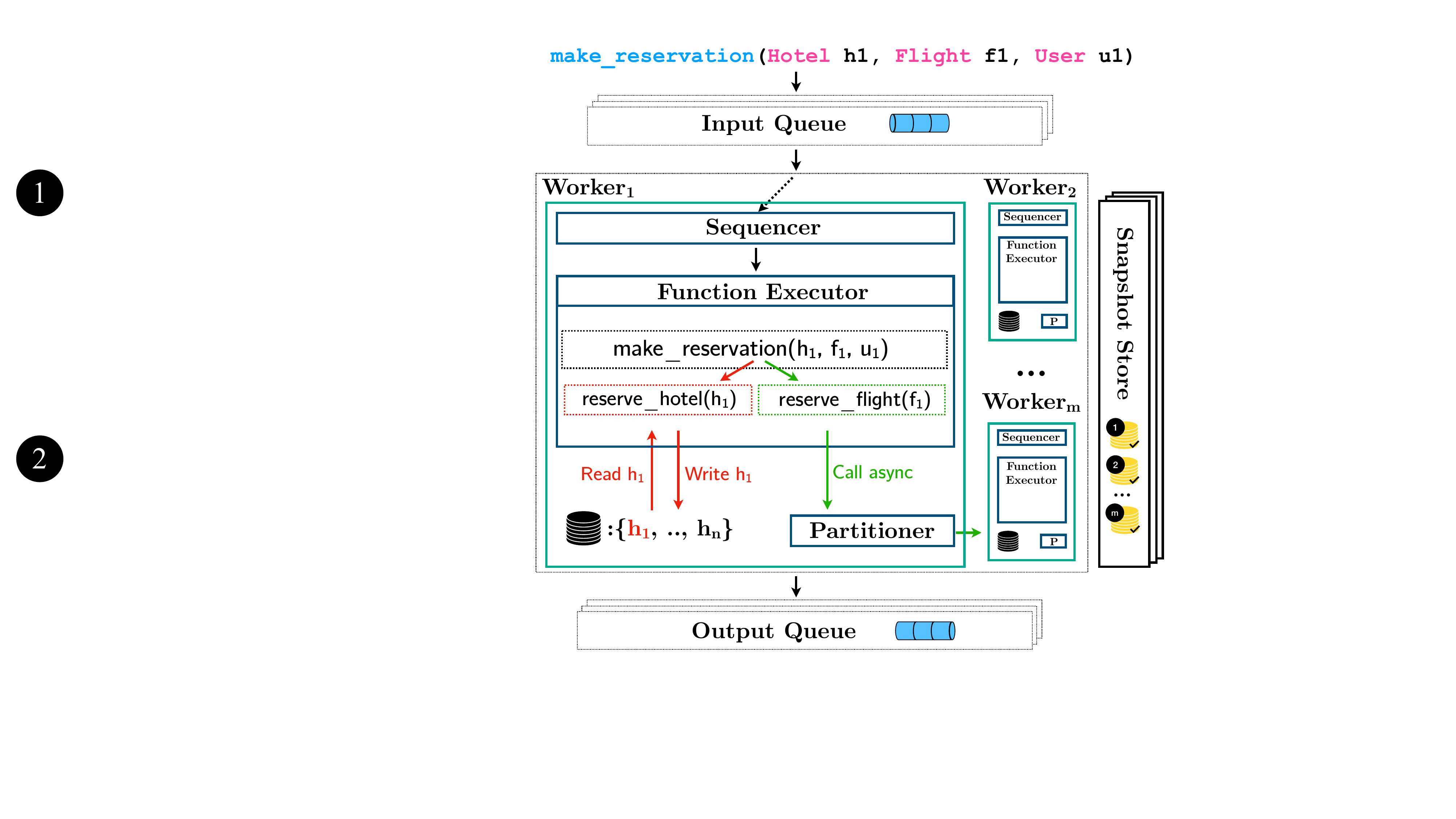}
        \caption{Stateful-Function execution in \sysname{}. In each worker, one coroutine manages the sequencing of incoming transactions, while another coroutine handles their processing. In this example, transaction (\texttt{make\_reservation}) consists of two functions: \texttt{reserve\_hotel} and \texttt{reserve\_flight}. A function can access local state (\texttt{reserve\_hotel}) but also perform remote calls to different partitions (\texttt{reserve\_flight}). This remote call uses the partitioner to locate the correct worker storing that partition.}
    \label{fig:styxfunction}
    \vspace{-4mm}
\end{figure}

\section{\sysname{}'s Architecture} 
\label{sec:dataflow-system}

In this section, we describe the components (\Cref{fig:styxfunction}) and the main design decisions of \sysname{}.

\subsection{Components}

\para{Coordinator} The coordinator manages and monitors \sysname{}'s workers, as well as the runtime state of the cluster (transactional metadata, dataflow state, partition locations, etc.). It also performs scheduling and health monitoring. \sysname{} monitors the cluster's health using a heartbeat mechanism and initiates the fault-tolerance mechanism (\Cref{sec:fault-tolerance}) once a worker fails.

\para{Worker} As depicted in \Cref{fig:styxfunction}, the worker is the primary component of \sysname{}, processing transactions, receiving or sending remote function calls, and managing state.
The worker consists of two primary coroutines. The first coroutine ingests messages for its assigned partitions from a durable queue and sequences them. The second coroutine receives a set of sequenced transactions and initiates the transaction processing. By utilizing the coroutine execution model, \sysname{} increases its efficiency since the most significant latency factor is waiting for network or state-access calls. Coroutines allow for single-threaded concurrent execution, switching between coroutines when one gets suspended during a network call, allowing others to make progress. Once the network call is completed, the suspended coroutine resumes processing.

\begin{figure*}[t]
    \centering
    \includegraphics[width=1.9\columnwidth]{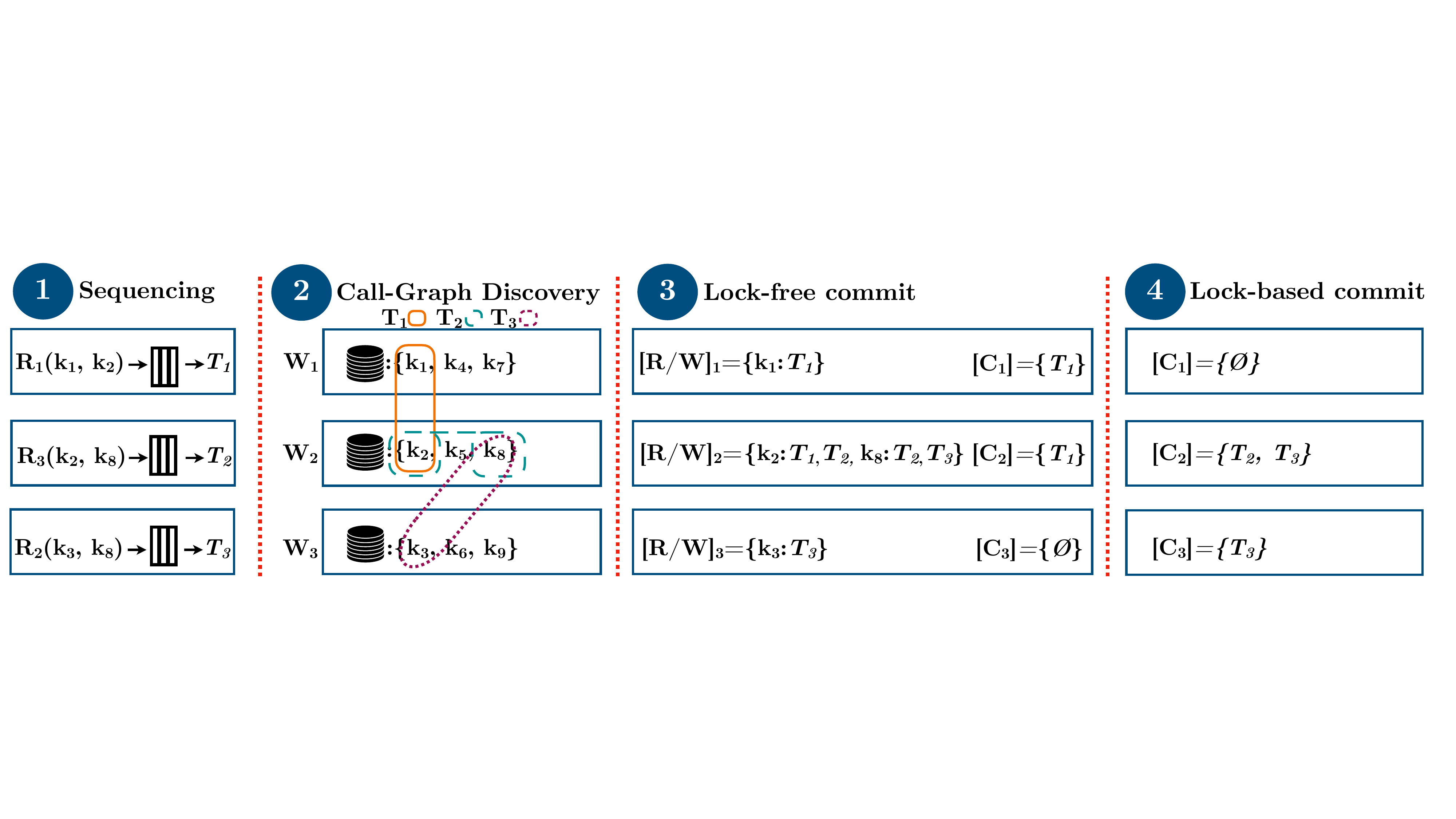}
        \caption{The transaction execution pipeline in \sysname{} is divided into $4$ parts. First, each external request ($R_i$) is sequenced as a transaction and is assigned a unique id. Afterward, the transactions execute their application logic, accessing local keys and performing remote function calls. While a transaction executes, \sysname{} tracks its accessed keys ($[R/W]_i$) and incrementally constructs its call-graph. Subsequently, \sysname{} commits the transactions that do not participate in unresolved conflicts without having to perform locking. For example, we observe that workers $W_{1}$ and $W_{2}$ are capable to commit $C_1 = C_2 = \bigl\{T_1\bigr\}$ while \textit{$T_{1}$} interacts with the same keys as \textit{$T_{2}$}; although it has the lowest id. In the final part, we commit all the transactions by resolving the conflicts with a lock-based mechanism ($C_2 = \bigl\{T_2, T_3\bigr\}$), $C_3 = \bigl\{T_3\bigr\}$).}
    \label{fig:styxlifecycle}
\end{figure*}

\para{Partitioning Stateful Entities Across Workers} \sysname{} makes use of the entities' \texttt{key} to distribute those entities and their state across a number of workers. By default, each worker is assigned a set of keys using hash partitioning.

\para{Input/Output Queue} For fault tolerance, \sysname{} assumes a persistent input queue from which it receives requests from external systems (e.g., from a REST gateway API). \sysname{} requires the input queue to be able to deterministically replay messages based on an offset when a failure occurs. As we detail in \Cref{sec:fault-tolerance}, the replayable input queue is necessary for \sysname{} to produce the same sequence of transactions after the recovery is complete and to enable early commit-replies (\Cref{sub:early}). In the same way, \sysname{} sends the result of a given transaction to an output queue from which an external system (e.g., the same REST gateway API) can receive it. Currently, \sysname{} leverages Apache Kafka \cite{kreps2011kafka}.

\para{Durable Snapshot Store} Alongside the replayable queue, durable storage is necessary for storing the workers' snapshots. Currently, \sysname{} uses Minio, an open-source S3 clone, to store the incremental snapshots as binary data files.

\subsection{Transaction Execution Pipeline}
\sysname{} employs an epoch-based transactional protocol that concurrently executes a batch of transactions in each epoch. A transaction may include multiple functions that, during runtime, form a call-graph of function invocations. Each function may mutate its entity's state, and the effects of function invocations are committed to the system state in a transactional manner. In \Cref{fig:styxfunction}, once \texttt{make\_reservation} enters the system, it is persisted and replicated by the input queue. Then, a worker ingests the call into its local sequencer that assigns a Transaction ID (TID) and processes all the encapsulated function calls as a single transaction. In the \texttt{make\_reservation} case, the transaction consists of two functions: \texttt{reserve\_hotel} and \texttt{reserve\_flight}. For this example, let us assume that \texttt{reserve\_hotel} is a local function call and \texttt{reserve\_flight} runs on a remote worker. \texttt{reserve\_hotel} will execute locally in an asynchronous fashion using coroutines and apply state changes. In contrast, \texttt{reserve\_flight} 
will execute asynchronously on a remote worker, applying changes on the remote state.

\begin{figure}[t]
\centering
    \includegraphics[width=0.9\columnwidth]{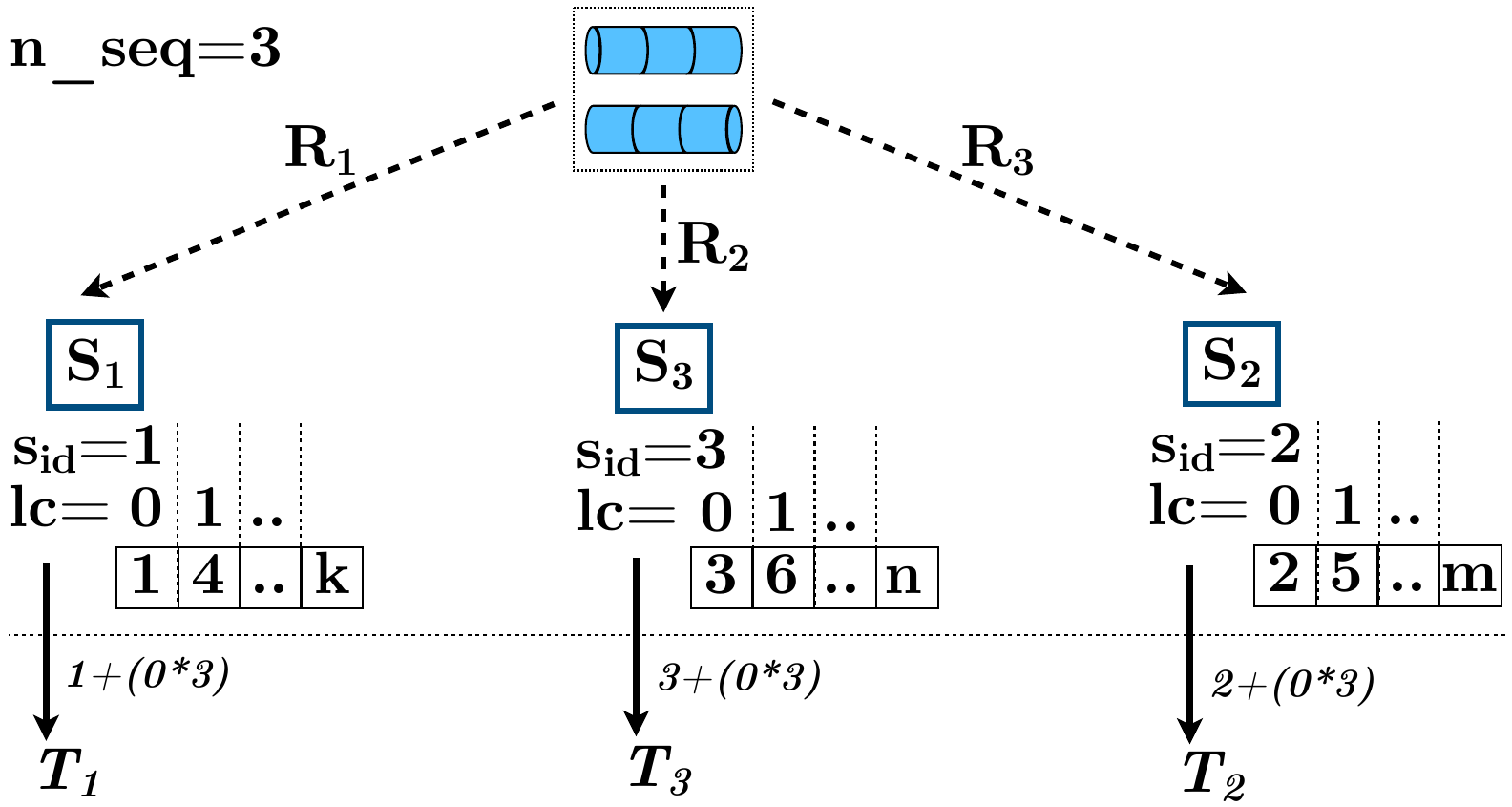}
        \caption{Example of TID assignment in \sysname{} with three sequencers. Their identifiers $\{1, 2, 3\}$ lead to the following sequences: $S_1 = \{1, 4, ..., k\}$,  $S_2 = \{2, 5, ..., m\}$,  $S_3 = \{3, 6, ..., n\}$ following the formula expressed in \Cref{eq:seq}.}
    \label{fig:sequencer}
\end{figure}

\section{Sequencing \& Function Execution} \label{sec:seq_f_ex}

The deterministic execution of functions with serializable guarantees requires a sequencing step that assigns a transaction ID (TID), which, in combination with the read/write (RW) sets, can be used for conflict resolution (\Cref{sec:cmt_t}). The challenge we tackle in this section is determining the boundaries of transactions (i.e., when a transaction's execution starts and finishes), which emerges from the execution of arbitrary function call-graphs \Cref{sec:ack_shares}.

\subsection{Transaction Sequencing} \label{sec:sequencing}
In this section, we discuss the sequencing mechanism (\textbf{\circled{1})} of \sysname{}. Deterministic databases ensure the serializable execution of transactions by forming a global sequence. In Calvin~\cite{thomson2012calvin}, the authors propose a partitioned sequencer that retrieves the global sequence by communicating across all partitions, performing a deterministic round-robin.

\para{Eliminating Sequencer Synchronization}
Instead of the original sequencer of Calvin that sends $\mathcal{O}(n^2)$ messages for the deterministic round-robin, \sysname{} adopts a method similar to the one followed by Mencius~\cite{barcelona2008mencius}, allowing \sysname{} to acquire a global sequence without any communication between the sequencers ($\mathcal{O}(1)$). This is achieved by having each sequencer assign unique transaction identifiers (TIDs) as follows:

\begin{equation} \label{eq:seq}
TID_{sid, lc} = sid + (lc * n\_seq)
\end{equation}

\noindent where $sid \in \mathbb{N}_1$ is the sequencer id assigned by the \sysname{} coordinator in the registration phase, $lc \in \mathbb{N}_0$ is a local counter of each sequencer specifying how many TIDs it has assigned thus far and $n\_seq \in \mathbb{N}_1$ is the total number of sequencers in the \sysname{} cluster. In the example of \Cref{fig:styxlifecycle}, the sequencers of the three workers will sequence $R_1$, $R_2$ and $R_3$ to $T_1$, $T_3$ and $T_2$ respectively. \Cref{fig:sequencer} illustrates how those TIDs are generated in parallel. Note that, conceptually, \sysname{} implements a partitioned sequencer where the global sequence $S = \{S_1 \cup S_2 \cup \dots \cup S_n\}$ is the union of all partitioned sequences.

\para{Mitigating Sequence Imbalance} In case a single sequencer $S_1$ receives more traffic than other sequencers, its local counter ($lc_1$) will increase more than the local counter of the rest of the sequencers. As a result, in the next epoch, sequencer $S_1$ would produce larger TIDs than the rest of the sequencers. This means that new transactions arriving at a less busy sequencer will receive higher priority for execution: transactions with higher TID receive less priority in our transactional protocol. In case of high contention in the workload, this would increase latencies for the busy ($S_1$) worker node. To avoid this, at the end of an epoch, the coordinator calculates the maximum $lc$ ($max(lc_1, lc_2, \ldots, lc_n)$) and communicates it to all workers so that they can adjust their local counter re-balancing sequences in every epoch. Balancing the workers' transaction priorities reduces the 99th percentile latency.

\para{Replication and Logging} There is no need to replicate and log the sequence within \sysname{} since the input is logged and replicated within the replayable queue. In case of failure, after transaction replay the sequencers will produce the exact same sequence (\cref{sec:seq_recovery}).


\subsection{Call-Graph Discovery}

After sequencing, \sysname{} needs to execute the sequenced transactions and determine their call-graphs and RW sets (\circled{2}). To this end, the function execution runtime ingests a given sequence of transactions to process in a given epoch. The number of transactions per epoch is either set by a time interval (by default, one millisecond) or by a configurable maximum number of transactions that can run per epoch (by default, 1000 transactions per epoch). We have chosen an epoch-based approach since processing the incoming transactions in batches increases throughput.

\sysname{}'s runtime executes all the sequenced transactions on a snapshot of the data to discover the read/write sets. Transactions that span multiple workers will implicitly change the read/write sets of the remote workers via function calls. There is an additional issue related to discovering the RW set of a transaction: before the functions execute, the call-graph of the transaction is unknown. This is an issue because the protocol requires all transactions to be completed before proceeding to the next phase. To tackle this problem, \sysname{} proposes a function acknowledgment scheme explained in more detail in \Cref{sec:ack_shares}.

After this phase, all the stateful functions that comprise transactions will have finished execution, and the RW sets will be known. In \Cref{fig:styxlifecycle}, transactions $T_1$, $T_2$, and $T_3$ will execute and create the following RW sets: $Worker_1 \rightarrow \{k_1: T_1\}$, $Worker_2 \rightarrow \{k_2: T_1,T_2$ and $k_8: T_2, T_3\}$ and $Worker_3 \rightarrow \{k_3: T_3\}$.

\begin{figure}[t]
\centering
    \includegraphics[width=\columnwidth]{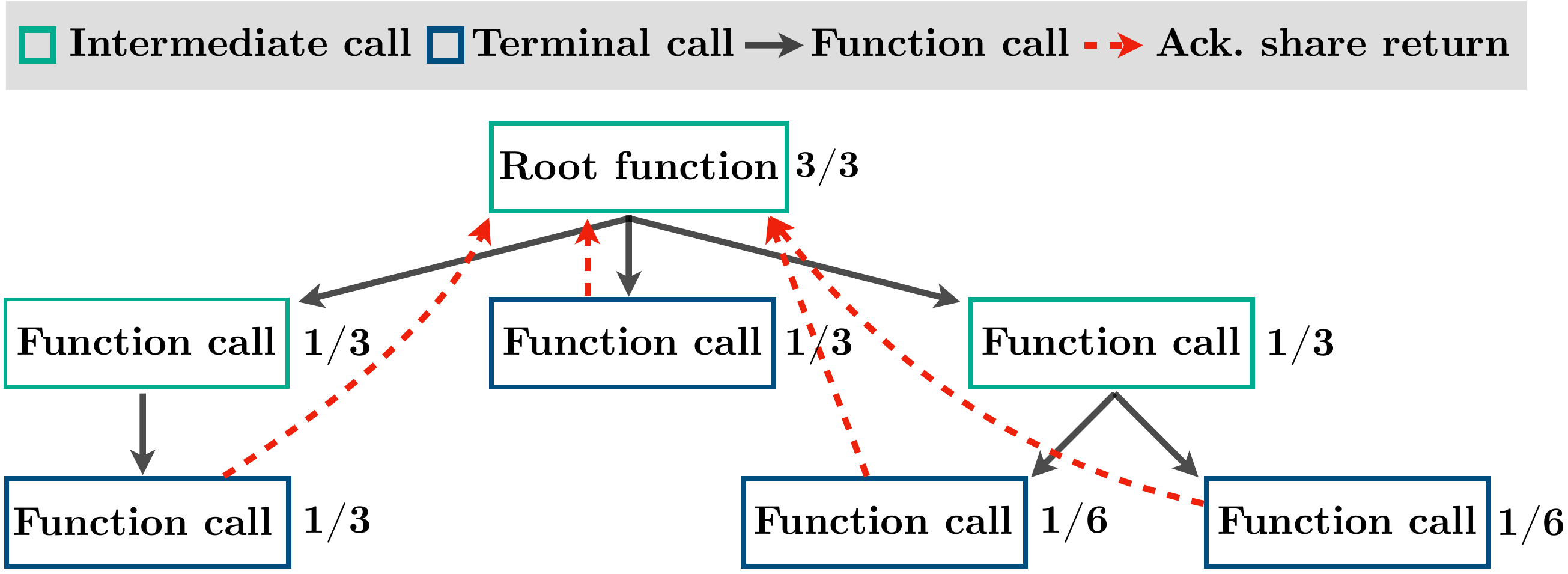}
        \caption{Asynchronous function call chains. A given root function call may invoke other functions throughout its execution. The original acknowledgment $(3/3)$ splits into parts as the function execution proceeds, and each function receives its own ack-share. For instance, in this function execution, the root function calls three other functions, thus splitting the ack-share into three equal parts. The same applies to subsequent calls, where the caller functions further split their ack-share. The sum of ack-shares of terminal (blue) calls (i.e., function calls that do not perform further calls) adds to exactly $3/3$, which allows the root function to report the completion of execution.}
    \label{fig:chains}
    \vspace{-5mm}
\end{figure}

\subsection{Function Execution Acknowledgment} \label{sec:ack_shares}
In the SFaaS paradigm, the call-graph formed by a transaction is unknown; functions could be coded by different developer teams and can form complex call-graphs. This uncertainty complicates determining when a transaction has completed processing, which is essential because phase \circled{3} can only start after all transactions have finished processing. To that end, each asynchronous function call of a given transaction is assigned an \texttt{ack\_share}. A given function knows how many shares to create by counting the number of asynchronous function calls during its runtime. The caller function then sends the respective acknowledgment shares to the downstream functions. For instance, in \Cref{fig:chains}, the transaction entry-point (root of the tree) calls three remote functions, splitting the ack\_share into three parts (3 x $1\//3$). The left-most function invokes only one other function and passes to it its complete ack\_share ($1\//3$). The middle function does not call any functions, so it returns the share to the root function when it completes execution, and the right-most function calls two other functions, splitting its share ($1\//3$) to 2 x $1\//6$. After all the function calls are complete, the root function should have collected all the shares. When the sum of the received shares adds to 1, the root/entry-point function can safely deduce that the execution of the complete transaction is complete.

This design is devised for two reasons: i)~if every participating function just sent an ack when it is done, the root would not know how many acks to expect in order to decide whether the entire execution has finished, and ii) if we used floats instead of fractions we could stumble upon a challenge related to adding floating point numbers.
For instance, if we consider floating-point numbers in the example mentioned above of the three function calls, the sum of all shares would not equal 1, but 0.99 since each share contributes 0.33. Subsequently, we cannot accurately round inexact division numbers; therefore, \sysname{} uses fraction mathematics instead.

A solution close to the \texttt{ack\_share} is the one of distributed futures \cite{distributedfutures}. However, it would not work in the SFaaS context as it either requires information about the entire call-graph for it to work asynchronously or it would need to create a chain of futures that would make it synchronous. Hence, it would introduce high latency for our use case.

\section{Committing Transactions} \label{sec:cmt_t}
After completing an epoch's call-graph discovery, \sysname{} needs to determine which transactions will commit and which will abort based on the transactions' Read/Write (RW) sets and TIDs. To this end, this section presents two different commit phases: $i)$ an optimistic lock-free phase that commits only the non-conflicting transactions, and $ii)$ a lock-based phase that only commits the transactions that were not able to commit in the first phase. The lock-based commit phase commits all conflicting transactions by acquiring locks in a TID-ordered sequence. To make the second phase faster, we have devised a caching scheme that can reuse the already discovered call-graph to avoid re-executing long function chains whenever possible (\cref{sec:caching}).

\subsection{Lock-free Commit Phase}

In case of conflict (i.e., a transaction $t$ writes a key that another transaction $t'$ also reads or writes on) similarly to~\cite{aria}, only the transaction with the lowest transaction ID will succeed to commit (\circled{3}). The transactions that have not been committed are put in a queue to be executed in the next phase \circled{4}  (maintaining their previously assigned ID).

In addition, workers ($W$) send their local conflicts to every other worker through the coordinator ($2*|W|$ messages): this way, every worker retains a global view of all the aborted/rescheduled transactions and can decide, locally, which transactions can be committed. Finally, note that transactions can also abort, not because of conflicts, but due to application logic causes (e.g., by throwing an exception). In that case, \sysname{} removes the related entries from the read/write sets to reduce possible conflicts further.

In this phase, all the transactions that have not been part of a conflict apply their writes to the state, commit, and reply to the clients. In the example shown in \Cref{fig:styxlifecycle}, only $T_1$ can commit in $W_1$ and $W_2$ due to conflicts in the RW sets of $W_2$ regarding $T_2$ and $T_3$; more specifically at keys $k_2$ and $k_8$.

\subsection{Lock-based Commit Phase}

In the previous phase, \circled{3}, only transactions without conflicts can be committed. We now explain how \sysname{} deals with transactions that have not been committed in a given epoch due to conflicts (\circled{4}). First, \sysname{} acquires locks in a given sequence ordered by transaction ID. Then, it reruns all transactions concurrently since all the read/write sets are known and commits them. However, if a transaction's read/write set changes in this phase, \sysname{} aborts the transaction and recomputes its read/write set in the next epoch. Now, in \Cref{fig:styxlifecycle}, $W_2$ can sequentially acquire locks for $T_2$ and $T_3$, leading to their commits in $W_2$ and $W_3$.

\subsection{Call-Graph Caching} \label{sec:caching}
As depicted in \Cref{fig:styxlifecycle}, the lock-based commit phase \circled{4} is used to execute any transactions that did not commit during the lock-free commit phase \circled{3}. \rev{By the time the lock-based commit phase starts, the state of the database may have changed since the lock-free commit. As a result, function invocations need to be re-executed to account for the data updates.} 

\begin{figure}[t]
\centering
    \includegraphics[width=\columnwidth]{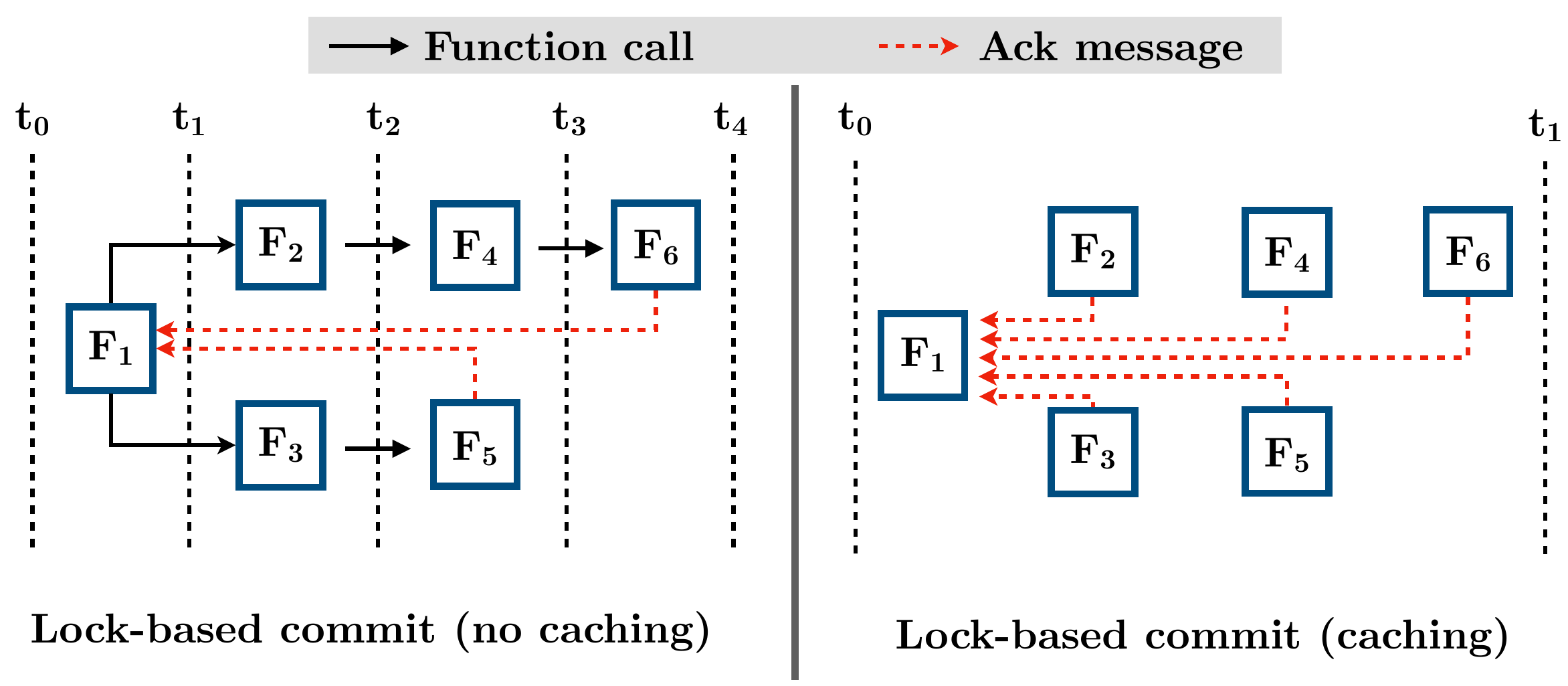}
        \vspace{-5mm}
        \caption{If no function caching is performed (left), the transaction execution will execute a deep call-graph; the messages will be sent sequentially and be equal to the number of function calls (5) in addition to the acks (2). \sysname{}'s function caching optimization (right) will lead to a concurrent function execution in the lock-based commit phase, between $t_0$ and $t_1$, and send only five acks asynchronously.}
    \label{fig:function_caching}
    \vspace{-9mm}
\end{figure}

On the left part of \Cref{fig:function_caching}, we depict such a function invocation. At time $t_0$, F$_1$ is invoked, which in turn invokes two function chains: $F_1 \rightarrow F_2 \rightarrow F_4 \rightarrow F_6$ and $F_1 \rightarrow F_3 \rightarrow F_5$. Once the two function chains finish their execution (on time $t_4$ and $t_3$ respectively), they can acknowledge their termination to the root call $F_1$. 

\para{Potential for Caching} During our early experiments, we noticed cases where $F_1$ is invoked and the parameters with which it calls $F_2$ (and in turn the invocations across the $F_1 \rightarrow ... \rightarrow F_6$ call chain) do not change. The same applied to the RW set of those function invocations; the RW sets remained unchanged. Since \sysname{} tracks those call parameters as well as the functions' RW sets, it can cache input parameters during the lock-free commit phase and reuse them during the lock-based commit, avoiding long sequential re-executions along the call chains. This case is depicted on the right part of \Cref{fig:function_caching}: the function-call chain does not need to be invoked in a sequential manner from $F_1$ all the way to $F_6$, leading to high latency. Instead, the individual workers can re-invoke those function calls locally and concurrently. As a result, all functions can execute in parallel and save on latency and network overhead ($t_4 - t_1$ in \Cref{fig:function_caching}).
\rev{Furthermore, caching does not require user input, is transparent to the API, and does not depend on the synchronous or asynchronous specification. Nonetheless, synchronous calls can be automatically transformed into asynchronous ones under certain conditions~\cite{beillahi2022automated, stateflow}.}

\para{Conditions for Parallel Function Re-invocation} Intuitively, if the parameters with which, e.g., $F_2$ is called, and the RW set of $F_2$ remains the same, we can safely assume that function $F_2$ can be invoked concurrently without having to be invoked sequentially by $F_1$. If those functions are successfully completed and acknowledge their completion to the root function $F_1$, it means that the transaction can be committed. To the contrary, if the RW set of any of the functions $F_1 - F_6$ changes, or the parameters of any of the functions along the call chains change, the transaction must be fully re-executed. In that case, \sysname{} will have to reschedule that transaction in the next epoch.

\subsection{Early Commit Replies via Determinism} \label{sub:early}

Implementing \sysname{} as a fully deterministic dataflow system offers a set of advantages involving the ability to communicate transaction commits to external systems (e.g., the client), even before the state snapshots are persisted to durable storage. A traditional transactional system can respond to the client only when $i)$ the requested transaction has been committed to a persistent, durable state or $ii)$~the write-ahead log is flushed and replicated. In \sysname{}'s case, that would mean when an asynchronous snapshot completes (i.e., is persisted to durable storage such as S3), leading to high latency.

Since \sysname{} implements a deterministic transactional protocol executing an agreed-upon sequence of transactions among the workers, after a failure, the system would run the same transactions with exactly the same effects. This determinism enables \sysname{} to give early commit replies: \emph{the client can receive the reply even before a persistent snapshot is stored.}
The assumption here is that the input queue, persisting the client requests, will provide to \sysname{}'s sequencers the requests in the same order after replay, a guarantee that is typically provided by most modern message brokers. 
Performing state mutations and message passing before persistence has also been explored in DARQ's speculative execution~\cite{darq}.





\section{Fault Tolerance}
\label{sec:fault-tolerance}




\sysname{} implements a coarse-grained fault tolerance mechanism. Instead of logging each function execution, it adopts a variant of existing checkpointing mechanisms used in streaming dataflow systems \cite{silvestre2021clonos,CarboneEF17,chandy1985distributed}. \sysname{} asynchronously snapshots state and stores it in a replicated fault-tolerant blob store (e.g., Minio / S3), enabling low-latency function execution. We describe \sysname{}'s fault tolerance mechanism below.

        \begin{algorithm}[t]
            \footnotesize
            \DontPrintSemicolon
            \SetAlgoLined
            \rev{
            \KwResult{Compacted Snapshot stored in durable storage}
            \SetKwInOut{Input}{Input}\SetKwInOut{Output}{Output}
            \SetKwComment{comm}{\hfill$\triangleright$\ }{}
            \Input{$\delta$: Delta changes, $O_{input}$: Input offset, $O_{output}$: Output offset, $E_{count}$: Epoch count, $SEQ_{count}$: Sequence count}
            \Output{$\mathcal{S}$: Compacted snapshot}
            \BlankLine
            
            \If{snapshotInterval}{
                state $\leftarrow \delta$ \comm*[r]{Prepare data and metadata for snapshot}
                metadata $\leftarrow \{O_{input}, O_{output}, E_{count}, SEQ_{count} \}$\;
                $\mathcal{S^{\delta}} \leftarrow$ serialize(state, metadata)\;
                store $\mathcal{S^{\delta}}$\; 
                inform coordinator\;
            }
            \If{compactionInterval}{
                $\mathcal{S} \leftarrow \emptyset$\;
                \ForEach{$\mathcal{S^{\delta}}$}{
                    $\mathcal{S} \leftarrow \text{compact}( \mathcal{S}, \mathcal{S^{\delta}})$ \comm*[r]{Compact delta snapshots}
                }
            }
            }
            \caption{\rev{Snapshotting Mechanism}}
            \label{algo:snapshot}
            \end{algorithm}

        \begin{algorithm}[t]
            \footnotesize
            \DontPrintSemicolon
            \SetAlgoLined
            \rev{
            \KwResult{Recovered state from durable storage, possible duplicate messages}
            \SetKwInOut{Input}{Input}\SetKwInOut{Output}{Output}
            \SetKwComment{comm}{\hfill$\triangleright$\ }{}
            \Input{
                $\mathcal{S}$: Latest compacted snapshot,\\
                $\mathcal{S^{\delta}}$: Incremental (delta) snapshots,\\
                $O^{last}_{output}$: Offset of last output,\\
            }
            \Output{
                $\mathcal{R}$: Set of possible duplicate messages, $state^s$: Snapshotted state, $O^s_{input}$: Snapshotted input offset, $O^s_{output}$: Snapshotted output offset, $ E^s_{count}$: Snapshotted epoch count, $SEQ^s_{count}$: Snapshotted sequencer count
            }
            \BlankLine
            \If{$\mathcal{S^{\delta}} \neq \emptyset$}{
                $\mathcal{S} \leftarrow \text{compact}(\mathcal{S},\mathcal{S^{\delta}})$ \comm*[r]{Compact delta snapshots, if any}
            }
            $state^s, O^s_{input}, O^s_{output}, E^s_{count}, SEQ^s_{count} \leftarrow$ deserialize $\mathcal{S'}$ \; \comm*[r]{Extract persisted state}
            $R \leftarrow \{ m \mid O^s_{\text{output}} \leq m \leq O^{last}_{\text{output}} \}$ \comm*[r]{Possible duplicates (\Cref{sec:exactly-once-output})}
            }
            \caption{\rev{Recovery Mechanism}}
            \label{algo:recovery}
            \end{algorithm}

\subsection{Incremental Snapshots \& Recovery} \label{sec:incr_sn_rec}

The snapshotting mechanism of \sysname{} resembles the approach of many streaming systems \cite{CarboneEF17, jet, SilvaZD16, ArmbrustDT18}, that extend the seminal Chandy-Lamport snapshots \cite{chandy1985distributed}. Modern stream processing systems checkpoint their state by receiving snapshot barriers at regular time intervals (epochs) decided by the coordinator. In contrast, \sysname{} leverages an important observation: workers do not need to wait for a barrier to enter the system in order to take a snapshot since the natural barrier in a transactional epoch-based system like \sysname{} is at the end of a transaction epoch.

\para{Snapshotting} To this end, instead of taking snapshots periodically by propagating markers across the system's operators, \sysname{} aligns snapshots with the completion of transaction epochs to take a consistent cut of the system's distributed state, including the state of the latest committed transactions, the offsets of the message broker, and the sequencer counters ($lc$).
\rev{The minimal information included in the snapshot is $O(N + c)$ where $N$ is the number of updates affecting the delta map, and $c$ is the fixed number of integers stored for the Kafka offsets and the sequencer variables.}

When the snapshot interval triggers, \sysname{} makes a copy of the current state changes to a parallel thread and persists incremental snapshots asynchronously, allowing \sysname{} to continue processing incoming transactions while the snapshot operation is performed in the background.
\rev{ The snapshotting procedure is described in Algorithm \ref{algo:snapshot}.}

\para{Recovery} In case of a system failure, \sysname{} $i)$ rolls back to the epoch of the latest completed snapshot, $ii)$ loads the snapshotted state, $iii)$ rolls back the replayable source's topic partitions (that are aligned with the \sysname{} operator partitions) to the offsets at the time of the snapshot, $iv)$ loads the sequencer counters, and finally, $v)$ verifies that the cluster is healthy before executing a new epoch. \rev{ The recovery procedure is described in Algorithm \ref{algo:recovery}.}

\begin{figure}[t]
\centering
    \includegraphics[width=0.85\columnwidth]{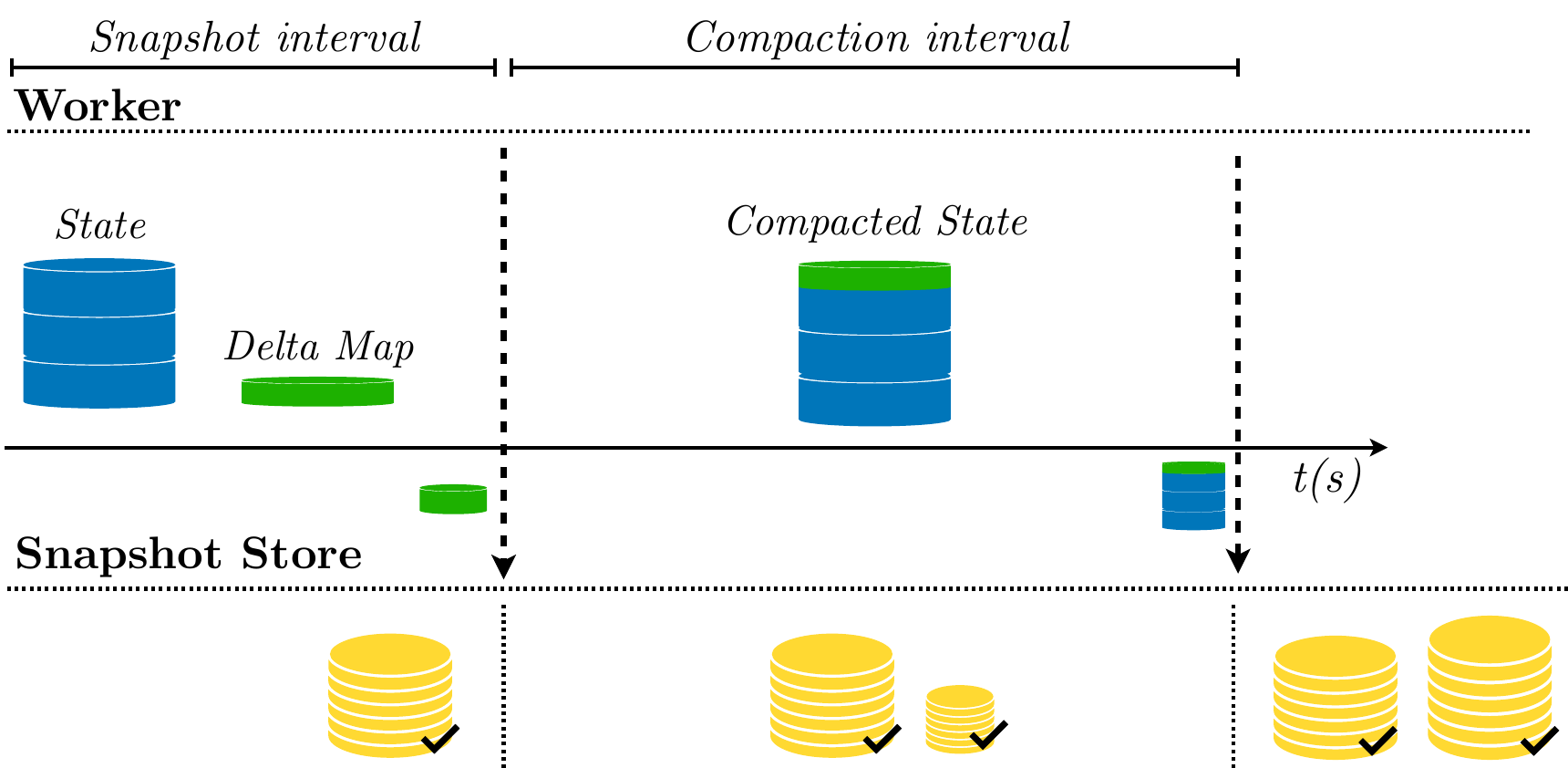}
        \caption{Incremental snapshots with Delta Maps in \sysname{}.}
    \label{fig:snapshot}
    \vspace{-5mm}
\end{figure}

\para{Incremental Snapshots \& Compaction} Each snapshot stores a collection of state changes in the form of \textit{delta maps}. A delta map is a hash table that tracks the changes in a worker's state in a given snapshot interval. When a snapshot is taken, only the delta map containing the state changes of the current interval is snapshotted. To avoid tracking changes across delta maps, \sysname{} periodically performs compactions where the deltas are merged in the background, as shown in \Cref{fig:snapshot}. \rev{The cost of compacting is equivalent to the cost of merging two hashmaps with the same key-spaces $(O(N))$. The total cost will be $O(M*N)$, with $M$ denoting the number of deltamaps we need to compact.}

\subsection{Sequencer Recovery} \label{sec:seq_recovery}
To guarantee determinism, upon recovery, \sysname{} 's sequencer needs to generate identical sequences as the ones generated between the latest snapshot and failure. The recovery protocol of the sequencer operates as follows: At first, during the snapshot, we store the local counter of each sequencer partition ($lc$) with its id ($sid$) and the epoch counter. Additionally, at the start of each epoch, \sysname{}  logs the number of transactions contained in that epoch, denoted as epoch size. Logging the epoch sizes is needed due to \sysname{} 's varying epoch sizes and the sequencer rebalancing scheme (\cref{sec:sequencing}). After failure, the recovered sequencer partitions are initialized with the snapshot's $lc$ and $sid$. Afterward, each partition retrieves from its log all the sizes of all epochs executed since the last snapshot. Finally, after recovery, the sequencer matches the epoch sizes to the ones recorded by the log, leading to the same global sequence observed before failure.

\subsection{Exactly-Once Processing}


At first, the durable input queue, which acts as a replayable source, allows \sysname{} to replay requests after failures. By rolling back the queue partitions (aligned with system operator partitions) to the respective offsets as recorded in the latest snapshot, \sysname{} can reprocess only the transactions whose state changes are not reflected yet in the snapshot. Transactions committed and early-commit replies stored in the egress can be deduplicated (\Cref{sec:exactly-once-output}).
 
\sysname{} runs each transaction to its completion in a single epoch. A given transaction can execute a large call-graph of functions that can affect the state. If a failure takes place, a transaction's state effects are restored to the latest snapshot, and the complete transaction is re-executed. As a result, no special attention is required to ensure that remote function calls are executed exactly-once, except for resetting all TCP channels between \sysname{}'s workers after recovery.


\rev{
\begin{lemma}\label{lem:exactlyonce}
The state mutations of committed transactions in \sysname{} are reflected exactly-once, even upon failure.
\end{lemma}

\begin{proof}

Let $S_t$ denote the state of the system at time $t$. $Q_t = \{r_1, \dots, r_n\}$ denotes the durable input queue at time $t$ that holds all requests $r_i$ to be processed. We assume that the input queue operates as FIFO and requests $r_i$ are deterministic.
Each $r_i$ will be sequenced as a transaction $T_i = \{upd_l, func_m\} $ by a deterministic sequencer, where $upd_l$ are the state updates and $func_m$ are the function calls of the transaction. We assume that $upd_l$  happens atomically and $func_m$ are also reflected once, given the use of a reliable communication protocol. Given the same initial state $S$ and input from $Q$, it always produces the same state transition $S \to S'$,which means $S'_{t+1} = mutation(S_t, Q_t)$. The execution of a transaction $T_i$ is deterministic.

At any time $t$, the state of the system $S_t$ reflects all transactions in $Q_t$ that have been fully executed and committed. Accordingly, the state $S_t$ ignores partially executed or in-progress transactions in $Q_t$. We denote the latest durable snapshot taken up to time $t$, as $\text{Snapshot}(S_t, i, n)$ where $n$ corresponds to the offsets of the first request $r_i$, and last request $r_n$ of the input queue to be processed up to time $t$. Upon failure, a subset of $Q_t$, $Q^{success}_t = \{r_1, \dots, r_k\}$ will contain successfully committed transactions and a subset $Q^{fail}_t = \{r_{k+1}, \dots, r_n\}$ will contain aborted transactions such that $Q_t = Q^{success}_t + Q^{fail}_t$. In order to recover from a failure, $Q_t$ is rolled back to $S_t$ from
$\text{Snapshot}(S_t, i, n)$ as we persist the offsets of our input queue. Transactions in $Q_t$ are replayed in the original order from offset $i$ to offset $n$ of our input queue. This is ensured by the FIFO queue and the deterministic sequencer. After processing the input transactions, $Q^{success}_t$ includes requests already reflected in $\text{Snapshot}(S_t)$, and $Q^{fail}_t$ includes pending requests. Since $\text{Snapshot}(S_t)$ reflects $Q^{success}_t$ and $Q_t = Q^{success}_t + Q^{fail}_t$, the replay and processing ensure: $S_{t+1}'' = mutation(S_t, Q^{fail}_t) = S_{t+1}'$. Thus, the effects of all transactions will be reflected in the state exactly-once, even after failure.
\end{proof}
}

\subsection{Exactly-Once Output} 
\label{sec:exactly-once-output}
A common challenge in the fault tolerance of streaming systems is that of the exactly once output~\cite{ElnozahyAW02, fragkoulis2024survey} in the presence of failures, which is hard to solve for low-latency use cases. For example, in Apache Flink's~\cite{flink} exactly-once output configuration, clients can only retrieve responses after those are persisted in a snapshot or a transactional sink. This arrangement is sufficient for streaming analytics but not for low-latency transactional workloads, as discussed previously in \Cref{sub:early}. 

To solve that, during recovery, \sysname{}: $i)$ reads the last offset of the egress topic, $ii)$ compares it with the output offset persisted in the snapshot, determining for which transactions the clients have already received replies, $iii)$ retrieves the TIDs attached in those replies, and $iv)$ does not send a reply again to the egress topic for those transactions. Note that this deduplication strategy is based on the fact that TIDs have been assigned deterministically.

\begin{figure}[t]
    \centering
    \includegraphics[width=0.75\columnwidth]{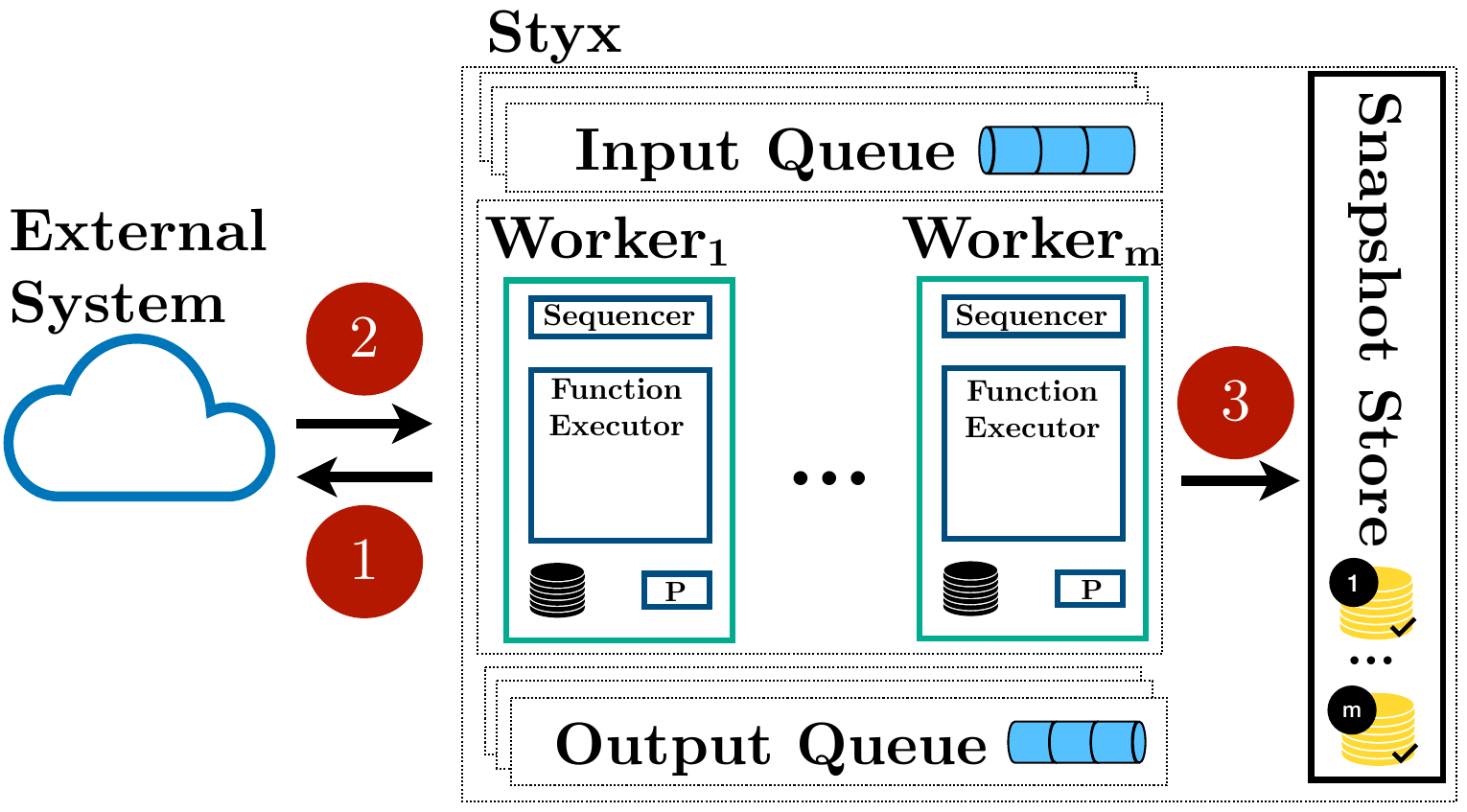}
    \vspace{-3mm}
        \caption{\rev{External system call critical points and \sysname{}.}}
    \label{fig:externalcalls}
\end{figure}

\rev{
\subsection{Addressing Non-Deterministic Functions} \label{sec:addr_non_det}
As discussed in \Cref{sec:incr_sn_rec} \sysname{}'s recovery mechanism is based on deterministic replay. To this end, \sysname{} requires that the functions authored by developers are also deterministic, i.e., replaying the same function multiple times, using the same inputs and database state, should yield the same results. However, one can achieve determinism even in the presence of non-deterministic logic inside functions, such as randomness (e.g., random numbers/sampling) or calls to external systems (e.g., calling an external database or API). \sysname{} can follow the approach of existing systems (e.g., Temporal~\cite{temporal}, Clonos~\cite{silvestre2021clonos}). In the following, we explain how this can be achieved.

\para{Randomness} To retain determinism in the case of randomness, \sysname{} can use an external fault tolerant write-ahead log (WAL) to log the random number along with the TID. Thus, in the case of failure and replay, \sysname{} can use the logged random number, essentially making the function call deterministic during replay.

\para{Calls to External Systems} As illustrated in \Cref{fig:externalcalls}, an interaction with an external system needs to consider three critical points to maintain determinism. \sysname{} assumes that the external system supports idempotency~\cite{ietf}, meaning that if a call is made twice with the same idempotency key, the effects on the external system's state and its return value will remain the same. In \circler{1} \sysname{} needs to log the idempotency key and the TID in the WAL before calling the external system. If the external system produces a response (\circler{2}), \sysname{} can store it in the WAL and retrieve it from there in case of replay. Finally, when \sysname{} completes a snapshot (\circler{3}), it can also clear the WAL for garbage collection since the prior entries are not needed.

Finally, \sysname{} could mask those operations behind an API that exposes the following functionality such as \texttt{system\_x.random} for random number generation  and \texttt{system\_x.call\_external} for external system calls.
}

\section{Evaluation} \label{sec:exp}

We evaluate \sysname{} by answering the following questions:
\begin{itemize}
  \renewcommand\labelitemi{--}
  \setlength\itemsep{0mm}
  \item (\Cref{sec:exp_lat_through}) How does \sysname{} compare to State-of-the-Art serializable transactional SFaaS systems? 
  \item (\Cref{sec:exp_skew}) How does \sysname{} perform under skewed workload? 
  \item (\Cref{sec:exp_scalability}) How well does \sysname{} scale? 
  \item (\Cref{sec:exp:snapshots}) Does the snapshotting mechanism affect performance? 
\end{itemize}

\subsection{Setup}
\label{sec:exp:setup}

\para{Systems Under Test} In the evaluation, we include SFaaS systems that provide serializable transactional guarantees. Those are:

\parait{Beldi~\cite{beldi}/Boki~\cite{boki}} Both systems use a variant of two-phase commit and Nightcore~\cite{nightcore} as their function runtime and store their data in DynamoDB. Additionally, Boki is deployed with the latest improvements of Halfmoon \cite{qi2023halfmoon}.

\parait{T-Statefun~\cite{tstatefun}} T-Statefun maintains the state and the coordination of the two-phase commit protocol within an Apache Flink cluster and ships the relevant state to remote stateless functions for execution. For fault tolerance, it relies on a RocksDB state backend that performs incremental snapshots. 

\parait{\sysname{}} \sysname{} is implemented in Python 3.12 and uses coroutines to enable asynchronous concurrent execution. Apache Kafka is used as an ingress/egress and Minio/S3 \cite{MinIO} as a remote persistent store for \sysname{}'s incremental snapshots. Finally, \sysname{} is a standalone containerized system that works on top of Docker and Kubernetes for ease of deployment.

\begin{figure*}[t]
    \centering
    \begin{subfigure}[t]{\columnwidth}
        \centering
        \includegraphics[width=\columnwidth]{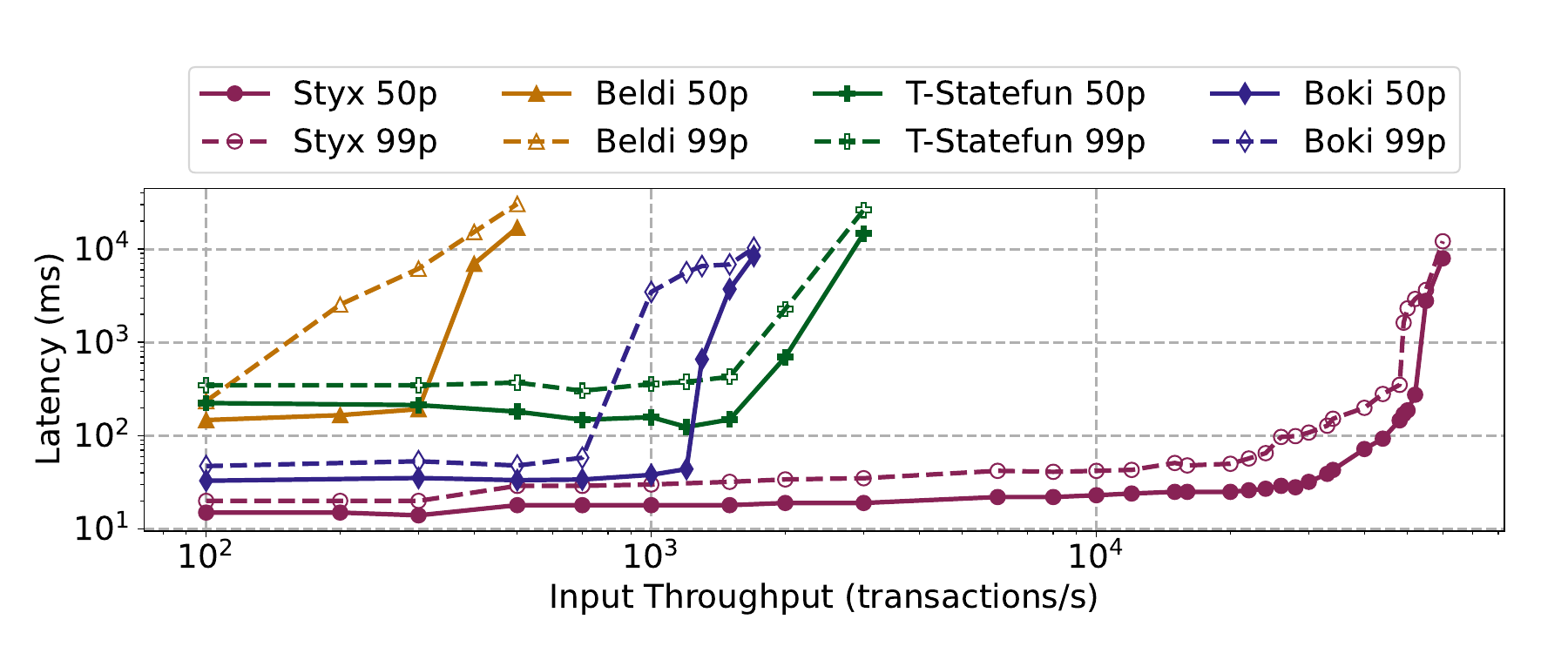}
        \vspace{-6mm}
        \caption[]%
        {{\footnotesize YCSB-T (uniform).}}
        \label{fig:tl_ycsbt}
    \end{subfigure}
    \hfill
    \begin{subfigure}[t]{\columnwidth}  
        \centering 
        \includegraphics[width=\columnwidth]{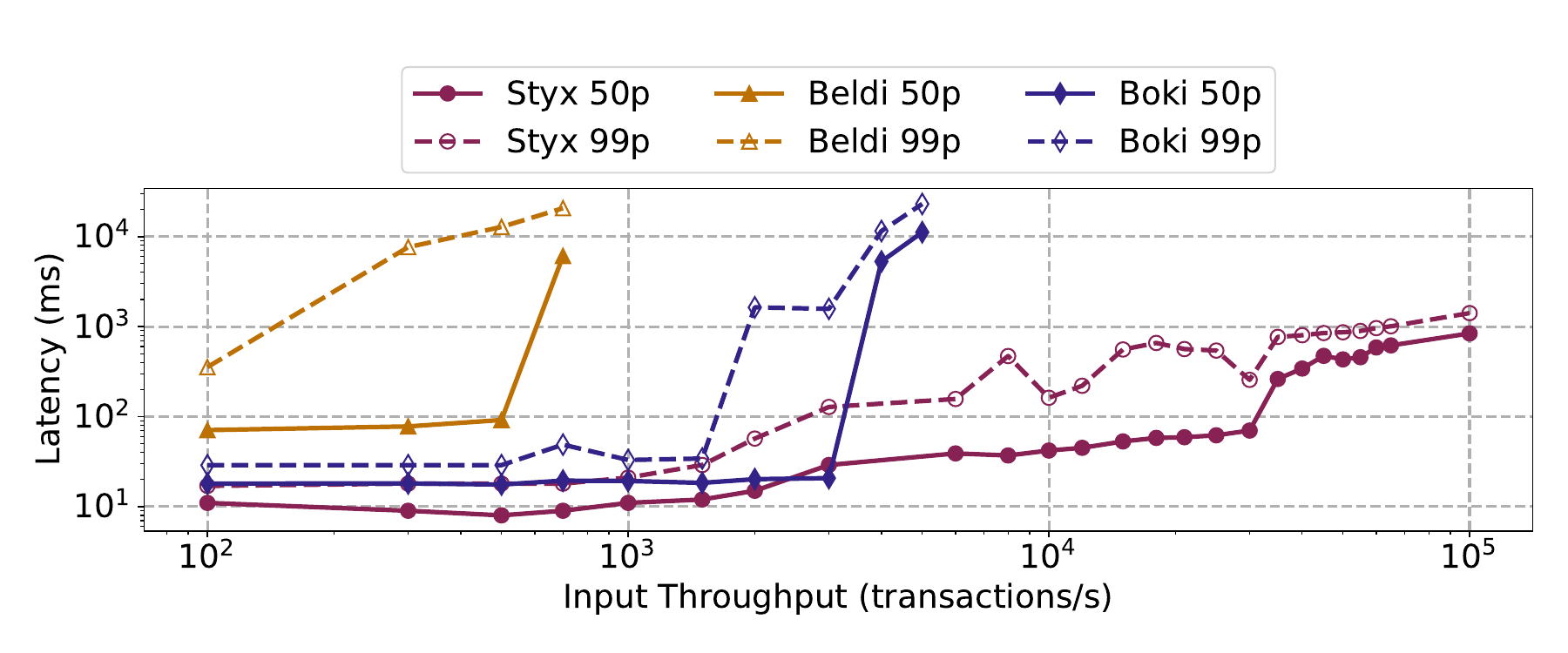}
                \vspace{-6mm}
        \caption[]%
        {{\footnotesize Deathstar Travel Reservation.}}    
        \label{fig:death_trav}
    \end{subfigure}
    \vskip\baselineskip
    \begin{subfigure}[t]{\columnwidth}   
        \centering 
        \includegraphics[width=\columnwidth]{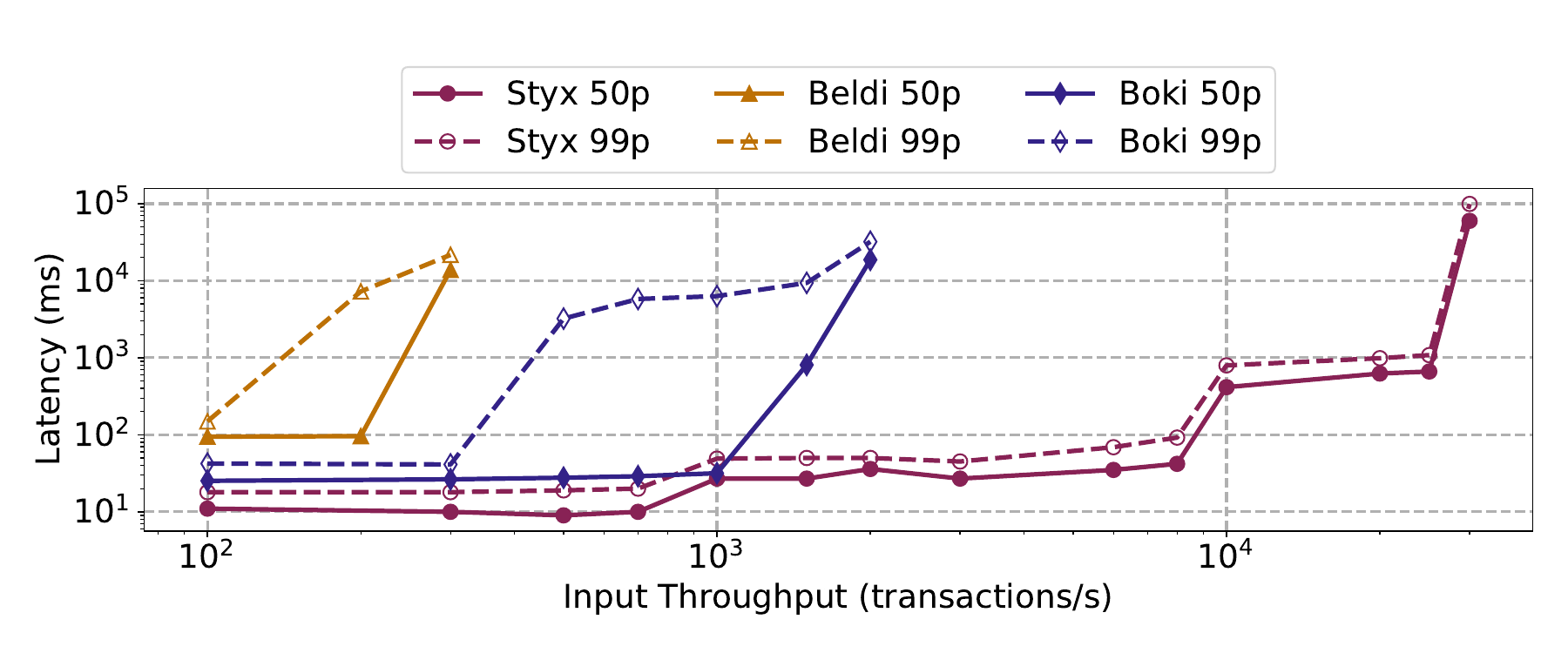}
                \vspace{-6mm}
        \caption[]%
        {{\footnotesize Deathstar Movie Review.}}    
        \label{fig:death_mov}
    \end{subfigure}
    \hfill
    \begin{subfigure}[t]{\columnwidth}   
        \centering 
        \includegraphics[width=\columnwidth]{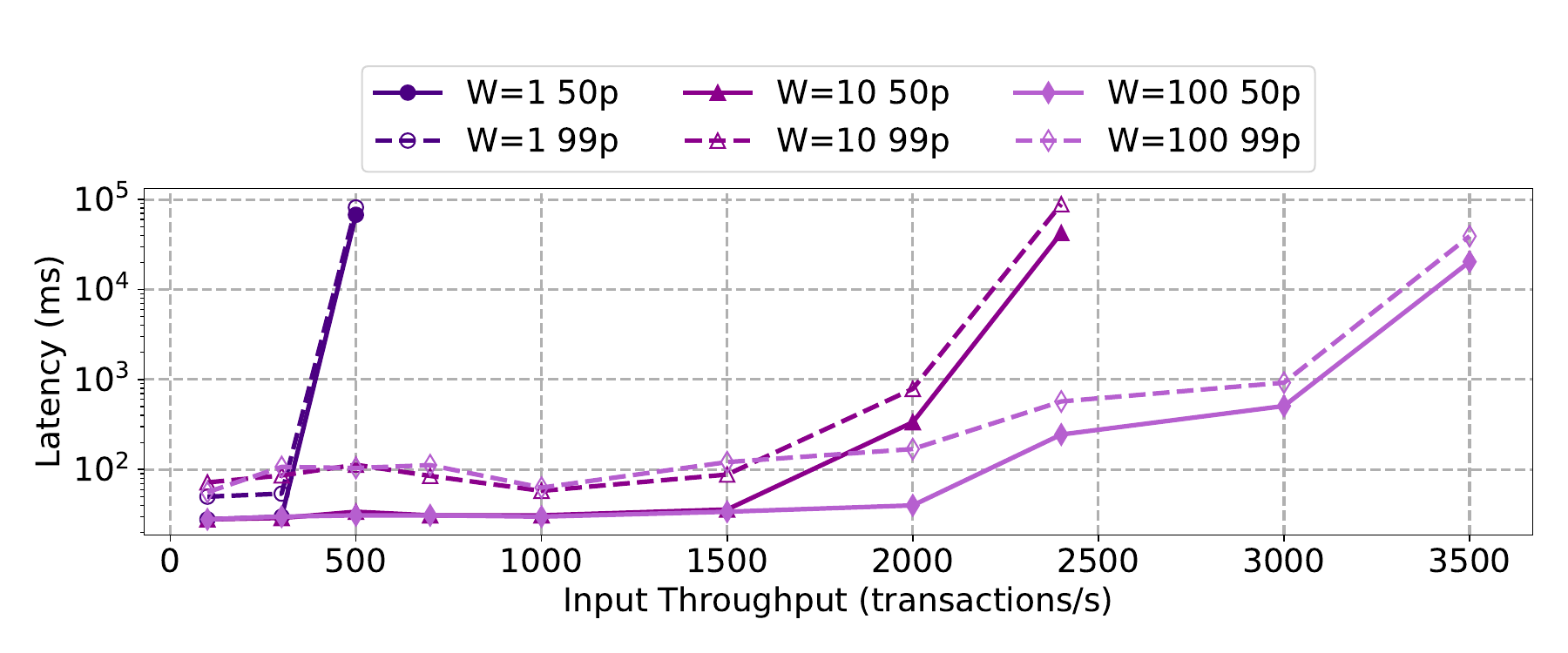}
        \vspace{-6mm}
        \caption[]%
        {{\footnotesize TPC-C on \sysname{} with 1, 10, and 100 warehouses.}}    
        \label{fig:styx_tpcc}
    \end{subfigure}
    \caption{Evaluation in different scenarios. T-Statefun does not support range queries required by the Deathstar workloads. TPC-C is only supported by \sysname{}.}
    \label{fig:throughput_latency}
\end{figure*}

\para{Workloads/Benchmarks} \Cref{tbl:scenaria} summarizes the three workloads used in the experiments. 

\begin{figure}[t]
\centering
  \includegraphics[width=\columnwidth]{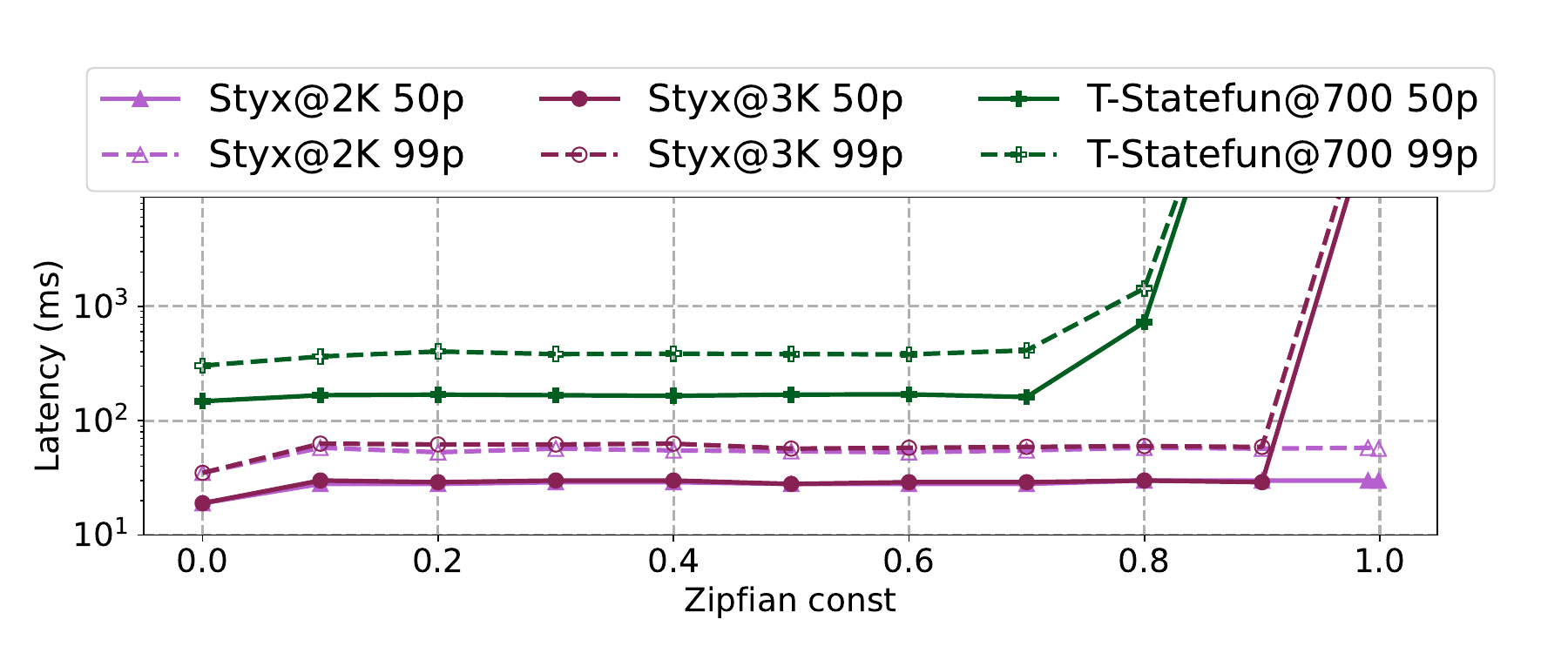} 
\caption{Latency evaluation for varying levels of contention (0.0 - 0.999) with YCSB-T (skewed). We ran \sysname{} with two different input throughput variations to show clearly its behavior under contention. Note that \sysname{} and T-Statefun execute all transactions to completion (abort\%=0).}
  \label{fig:zipf_styx_ts}
\end{figure}

\parait{YSCB-T~\cite{dey2014ycsb+}} We use a variant of YCSB-T~\cite{dey2014ycsb+}  where each transaction consists of two reads and two writes. The concrete scenario is as follows: First, we create 10.000 bank accounts (keys) and perform transactions in which a debtor attempts to transfer credit to a creditor. This transfer is subject to a check on whether the debtor has sufficient credit to fulfill the payment. If not, a rollback needs to be performed. The selection of a relatively small number of keys is deliberate: we want to assess the systems' ability to sustain transactions under high contention. In addition, for the experiment depicted in \Cref{fig:zipf_styx_ts} (skewed distribution), we select the debtor key based on a uniform distribution and the creditor based on a Zipfian distribution, where we can vary the level of contention by modifying the Zipfian coefficient.

\parait{Deathstar~\cite{deathstar}} We employ Deathstar~\cite{deathstar}, as adapted to SFaaS workloads by the authors of Beldi~\cite{beldi}. It consists of two workloads: $i)$~the Movie workload implements a movie review service where users write reviews about movies. $ii)$~the Travel workload implements a travel reservation service where users search for hotels and flights, sort them by price/distance/rate, find recommendations, and transactionally reserve hotel rooms and flights. Both Deathstar workloads follow a uniform distribution. Note that T-Statefun could not run in this set of experiments since it does not support range queries.

\parait{TPC-C~\cite{tpcc}} The prime transactional benchmark targeting OLTP systems is TPC-C~\cite{tpcc}. In our evaluation, we employ the NewOrder and Payment transactions, and we had to rewrite them into the SFaaS paradigm, splitting the NewOrder transaction into 20-50 function calls (one call for each item in the NewOrder transaction) and the Payment transaction into 8 function calls. TPC-C scales in size/partitions by increasing the number of warehouses represented in the benchmark. While a single warehouse represents a skewed workload (all transactions will hit the same warehouse), increasing the number of warehouses decreases the contention, allowing for higher throughput and lower latency. Note that the TPC-C experiments do not include Beldi, Boki, or T-Statefun because they do not support it.

\begin{table}[t]
\centering
\resizebox{\columnwidth}{!}{
    \begin{tabular}{l||c|c|c}
    \textbf{Scenario} & \textbf{\#keys} & \textbf{Function Calls} & \textbf{Transactions} \% \\ \hline \hline
    \textbf{YCSB-T} & 10k & 2 & 100\% \\ \hline
    \textbf{Deathstar Movie} & 2k & 9-10 & 0\% \\ \hline
    \textbf{Deathstar Travel} & 2k & 3 & 0.5\% \\ \hline
    \textbf{TPC-C} & 1m-100m & 8 / 20-50 & 100\%
    \end{tabular}
    }
    \vspace{2mm}
    \caption{Workload characteristics.}
    \label{tbl:scenaria}
    \vspace{-8mm}
\end{table}

 \para{Resources} For Beldi/Boki, T-Statefun and \sysname{}, we assigned a total of 112 CPUs with 2GBs of RAM per CPU, matching what is presented in the original Boki paper \cite{boki}.  \rev{Additionally, throughout all the evaluation scenarios, the data fit in memory across all systems.} Unless stated otherwise, \sysname{} and T-Statefun are configured to perform incremental snapshots every 10 seconds. All external systems, i.e., DynamoDB (Beldi, Boki),  Minio, and Kafka (\sysname{}, T-Statefun), are configured with three replicas for fault tolerance.

\parait{External Systems} Boki and Beldi use a fully managed DynamoDB instance at AWS, which does not state the amount of resources it occupies and is additional to the 112 CPUs assigned to Boki and Beldi. Similarly, the resources assigned to Minio/S3 (\sysname{} and T-Statefun) are not accounted for.

\para{Metrics} Our goal is to observe systems' behavior, measured by their latency while varying the input throughput. 

\noindent\underline{\textit{Input throughput}} represents the number of transactions submitted per second to the system under test. As the input throughput increases during an experiment, we expect the latency of individual transactions to increase until aborts start to manifest due to contention or high load.

\noindent\underline{\textit{Latency}} represents the time interval between submitting a transaction and the reported time when the transaction is committed/aborted. In \sysname{} and T-Statefun, the latency timer starts when a transaction is submitted in the input queue (Kafka) and stops when the system reports the transaction as committed/aborted in the output queue. Similarly, in Beldi and Boki, the latency is the time since the input gateway has received a transaction and the time that the gateway reports that the transaction has been committed/aborted.

\subsection{Latency vs. Throughput} \label{sec:exp_lat_through} \label{sec:exp_skew}

We first study the latency-throughput tradeoff of all systems. We retain the resources given to the systems constant (112 CPUs) while progressively increasing the input throughput. We measure the transaction latency. As depicted in \Cref{fig:throughput_latency}, \sysname{} outperforms its baseline systems by at least an order of magnitude. Specifically, in YCSB-T (\Cref{fig:tl_ycsbt}), \sysname{} achieves a performance improvement of \textasciitilde20x in terms of throughput against T-Statefun, which ranks second. In addition, \sysname{} outperforms Boki by \textasciitilde30x in Deathstar's travel reservation workload (\Cref{fig:death_trav}) and  by \textasciitilde35x in Deathstar's movie review \Cref{fig:death_mov}) workload. Finally, in the TPC-C benchmark (\Cref{fig:styx_tpcc}), which requires a large number of function calls per transaction (20-50), we observe that \sysname{}'s performance improves as we increase the input throughput for different numbers of warehouses, reaching up to 3K TPS with sub-second 99$^{\text{th}}$ percentile latency (100 warehouses).


\para{Aborts \& Throughput} Beldi and Boki follow a no-wait-die concurrency control approach, which leads to a significant amount of aborts as the throughput increases. \sysname{} and T-Statefun do not use such a transaction abort mechanism. Instead, they execute all transactions to completion. This difference in handling transactions under high load makes the latencies across systems hard to compare. For this reason, in \Cref{fig:zipf_styx_ts}, we plot the results of \sysname{} and T-Statefun and present the performance of Beldi and Boki in a separate table (\Cref{fig:bbabort}), alongside their abort rates.

\begin{table}[t]
\resizebox{\columnwidth}{!}{
\begin{tabular}{rccccccccc}
\multicolumn{1}{l}{} & \multicolumn{1}{l}{} & 0.0 & 0.2 & 0.4 & 0.6 & 0.8 & 0.9 & 0.99 & 0.999 \\ \hline \hline
\multicolumn{1}{r|}{Beldi} & \multicolumn{1}{c||}{Abort \%} & \multicolumn{1}{c|}{47.93} & \multicolumn{1}{c|}{45.54} & \multicolumn{1}{c|}{44.31} & \multicolumn{1}{c|}{47.28} & \multicolumn{1}{c|}{52.40} & \multicolumn{1}{c|}{56.06} & \multicolumn{1}{c|}{61.62} & \multicolumn{1}{c|}{60.70} \\
\multicolumn{1}{r|}{} & \multicolumn{1}{c||}{CMT TPS} & \multicolumn{1}{c|}{104} & \multicolumn{1}{c|}{108} & \multicolumn{1}{c|}{111} & \multicolumn{1}{c|}{105} & \multicolumn{1}{c|}{95} & \multicolumn{1}{c|}{76} & \multicolumn{1}{c|}{76} & \multicolumn{1}{c|}{78} \\ \hline
\multicolumn{1}{r|}{Boki} & \multicolumn{1}{c||}{Abort \%} & \multicolumn{1}{c|}{48.77} & \multicolumn{1}{c|}{48.23} & \multicolumn{1}{c|}{49.54} & \multicolumn{1}{c|}{51.82} & \multicolumn{1}{c|}{61.29} & \multicolumn{1}{c|}{68.50} & \multicolumn{1}{c|}{74.47} & \multicolumn{1}{c|}{70.71} \\
\multicolumn{1}{r|}{} & \multicolumn{1}{c||}{CMT TPS} & \multicolumn{1}{c|}{359} & \multicolumn{1}{c|}{362} & \multicolumn{1}{c|}{353} & \multicolumn{1}{c|}{337} & \multicolumn{1}{c|}{271} & \multicolumn{1}{c|}{220} & \multicolumn{1}{c|}{179} & \multicolumn{1}{c|}{205} \\ \hline
\end{tabular}
}
    \vspace{2mm}
    \caption{Evaluation of Boki and Beldi for varying levels of contention with YCSB-T. We report the abort ratio and committed transactions rate and omit latency since the systems do not execute all transactions to completion. Both run at their maximum sustainable throughput.}
    \vspace{-8mm}
    \label{fig:bbabort}
\end{table}

We observe the following: $i)$ at the highest level of contention ($Zipfian$ at $0.999$) \sysname{} achieves at least 2000 TPS, outperforming the rest by \textasciitilde 5-10x in terms of effective throughput, $ii)$ both Beldi and Boki (that run at their maximum sustainable throughput) abort more transactions as the level of contention increases (\textasciitilde40-70\%), which significantly impacts their effectiveness as shown in \Cref{fig:bbabort}, and $iii)$ \sysname{} shows an increase in latency only in high levels of contention ($Zipfian>0.99$) while executing at \textasciitilde4x higher throughput than the rest.

\para{Runtime Breakdown} In \Cref{fig:breakdown}, we show where the systems under test spend their processing time. We use YCSB-T for this purpose since it is the only benchmark supported by all the systems (\Cref{sec:exp:setup}). We measured the median latency while all the systems were running at 100 TPS for 60 seconds and averaged the proportions of function execution, networking, and state access across all committed transactions. The key observations are: $i)$ \sysname{}'s co-location of processing and state led to minimal state access latency, and $ii)$ \sysname{}'s asynchronous networking allows for lower network latency.

\begin{table}[t]
\resizebox{1\columnwidth}{!}{
\begin{tabular}{c||c|c|c}
\textbf{System} & \textbf{Function Execution} & \textbf{Networking} & \textbf{State Access} \\ \hline \hline
\textbf{\sysname{}} & \textbf{0.34ms} - 2.2\% & \textbf{14.33ms} - 95.6\% & \textbf{0.32ms} - 2.2\% \\ \hline
\textbf{Boki} & 1.1ms - 3.3\% & 16.1ms - 49\% & 15.68ms - 47.7\% \\ \hline
\textbf{T-Statefun} & 2.76ms - 2.2\% & 92.12ms - 74.3\% & 29.11ms - 23.5\% \\ \hline
\textbf{Beldi} & 1.01ms - 0.7\% & 56.58ms - 38.4\%& 89.57ms - 60.9\% \\ \hline
\end{tabular}
}
    \vspace{2mm}
    \caption{Performance breakdown of all systems. (median latency - percentage from the total)}
    \label{fig:breakdown}
    \vspace{-8mm}
\end{table}

\para{Takeaway} The rather large performance advantages of \sysname{} across all experiments are enabled by the following three properties and design choices: $i)$ the co-location of processing and state with efficient networking as shown in \Cref{fig:breakdown}, contrary to the other systems that have to transfer the state to their function execution engines; $ii)$ the asynchronous snapshots with delta maps for fault tolerance compared to the replication of Beldi/Boki and the LSM-tree-based incremental snapshots of T-Statefun; $iii)$ the efficient transaction execution protocol employed in \sysname{} compared to the two-phase commit used by \sysname{}'s competition.

\subsection{Scalability} 
\label{sec:exp_scalability}

\begin{figure}[t]
    \centering
    \includegraphics[width=\columnwidth]{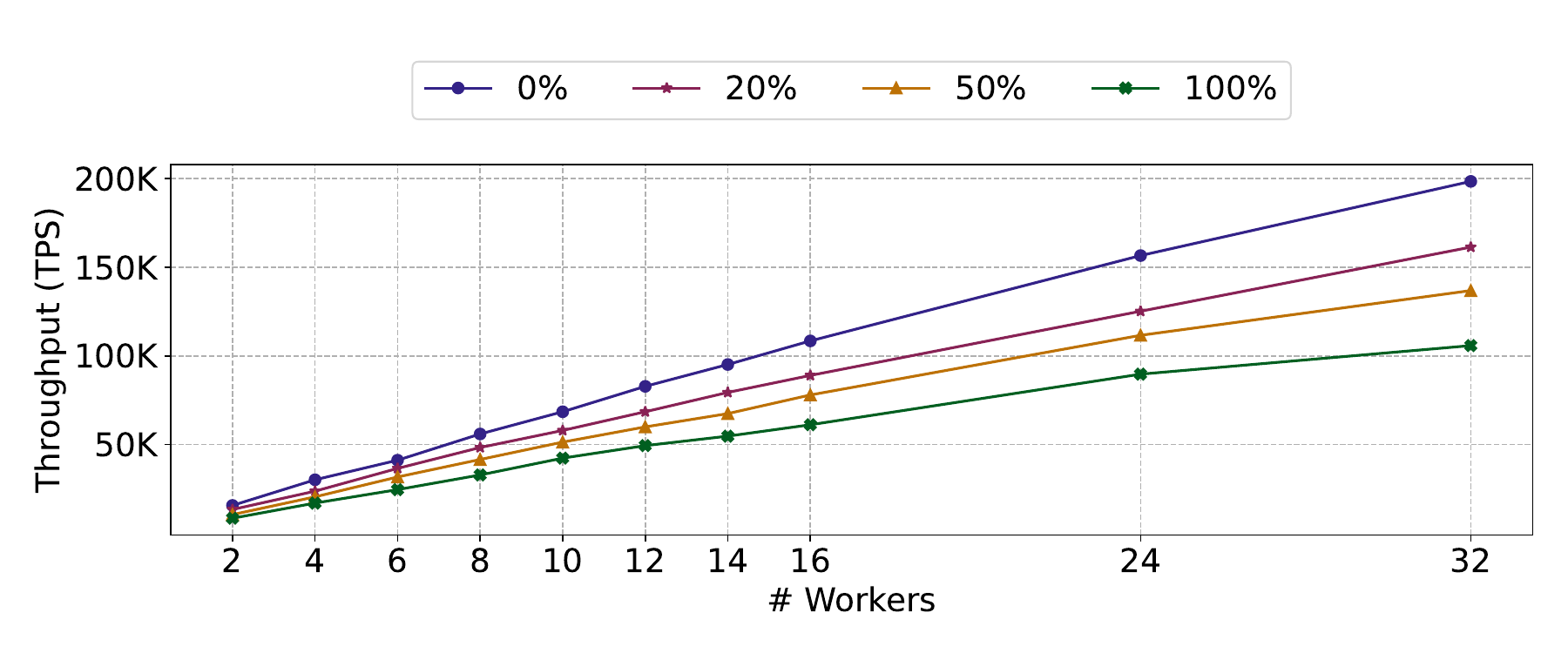}
    \vspace{-6mm}
    \caption{Scalability of \sysname{} on YCSB-T with varying percentages of multi-partition transactions.}
    \label{fig:scalability}
    \vspace{-5mm}
\end{figure}

In this experiment, we test the scalability of \sysname{} by increasing the number of \sysname{} workers. Each worker is assigned 1 CPU each and a state of 1 million keys. We measure the maximum throughput on YCSB-T. The goal is to calculate the speedup of operations as the input throughput and number of workers scale together. In addition, we control the percentage of multi-partition transactions in the workload, i.e., transactions that span across workers. In \Cref{fig:scalability}, we observe that in all settings, \sysname{} retains near-linear scalability. Finally, \sysname{} displays the expected behavior as the number of multi-partition transactions increases. 

\subsection{Fault-Tolerance Evaluation}
\label{sec:exp:snapshots}

\begin{figure}[t]
     \centering
     \begin{subfigure}[b]{0.49\columnwidth}
        \centering
        \includegraphics[width=\columnwidth]{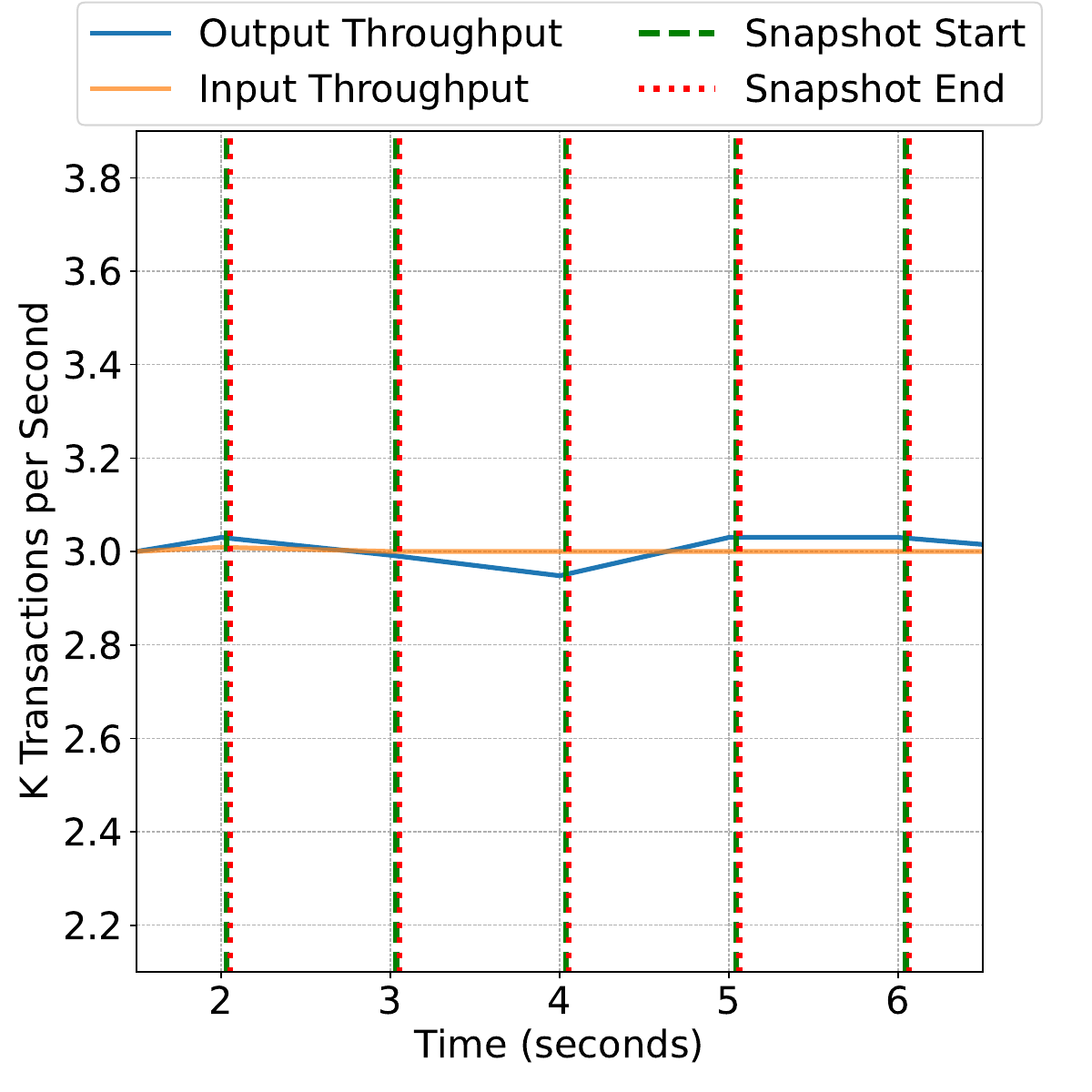}
            \caption{Throughput}
        \label{fig:exp-t-snap}
     \end{subfigure}
     \begin{subfigure}[b]{0.49\columnwidth}
        \centering
        \includegraphics[width=\columnwidth]{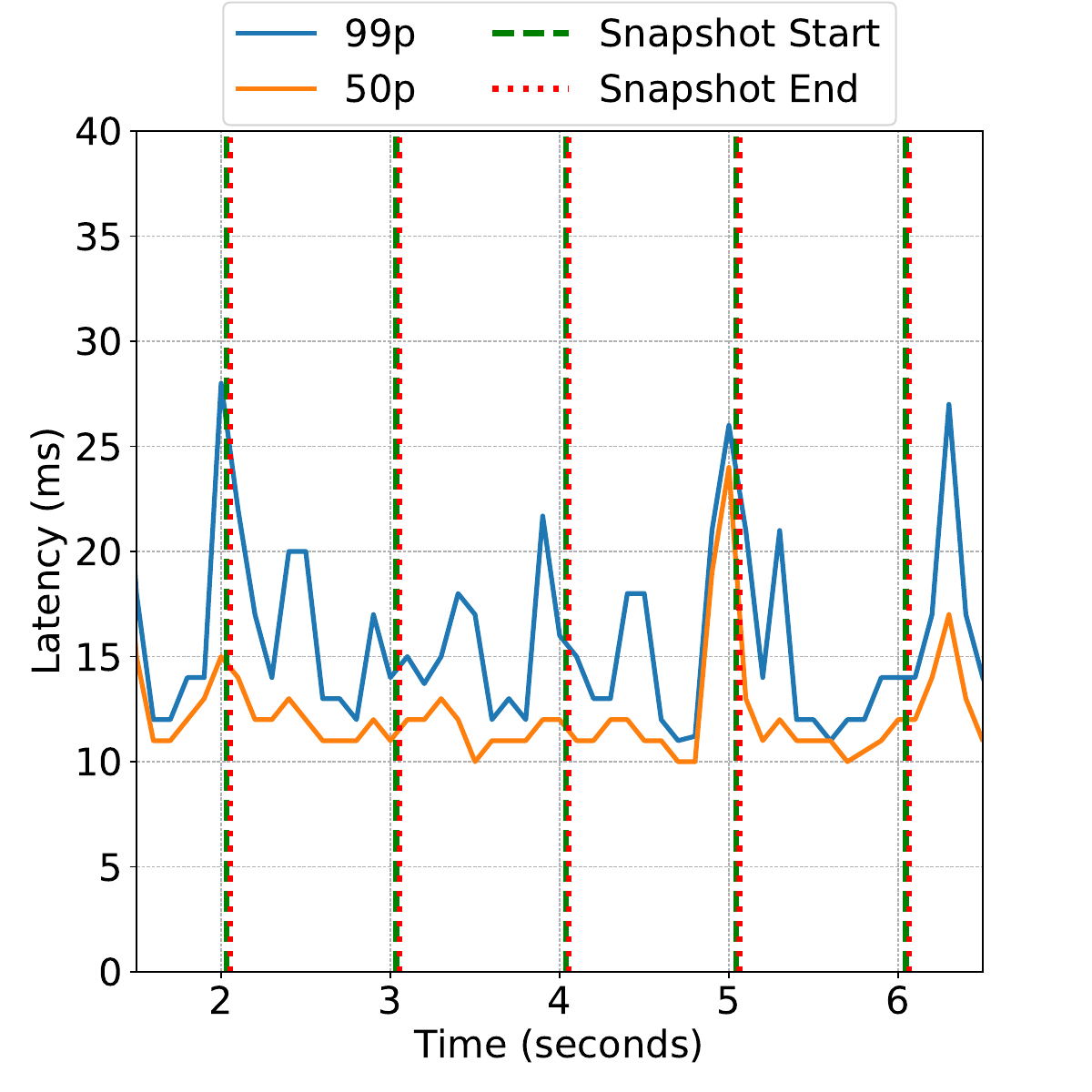}
            \caption{Latency}
        \label{fig:exp-l-snap}
     \end{subfigure}
        \caption{Impact of \sysname{}'s snapshotting on performance}
        \label{fig:exp-snapshots}
\end{figure}

\para{Effect of Snapshots} In \Cref{fig:exp-snapshots}, we depict the impact of the asynchronous incremental snapshots to \sysname{}'s performance. In both figures, we mark when a snapshot starts and ends. The state includes 1 million keys, and we use a 1-second snapshot interval. \sysname{} is deployed with four 1-CPU workers, and the input transaction arrival rate is fixed to 3K YCSB-T TPS. In \Cref{fig:exp-t-snap}, we observe that during a snapshot operation, \sysname{} shows virtually no performance degradation in throughput. In \Cref{fig:exp-l-snap}, we observe a minor increase in the end-to-end latency in some snapshots. The reason for that is the concurrent snapshotting thread, which competes with the transaction execution thread during snapshotting. At the same time, it also has to block the transaction execution thread momentarily to copy the corresponding operator's state delta.

\begin{figure}[t]
     \centering
     \begin{subfigure}[b]{0.49\columnwidth}
        \centering
        \includegraphics[width=\columnwidth]{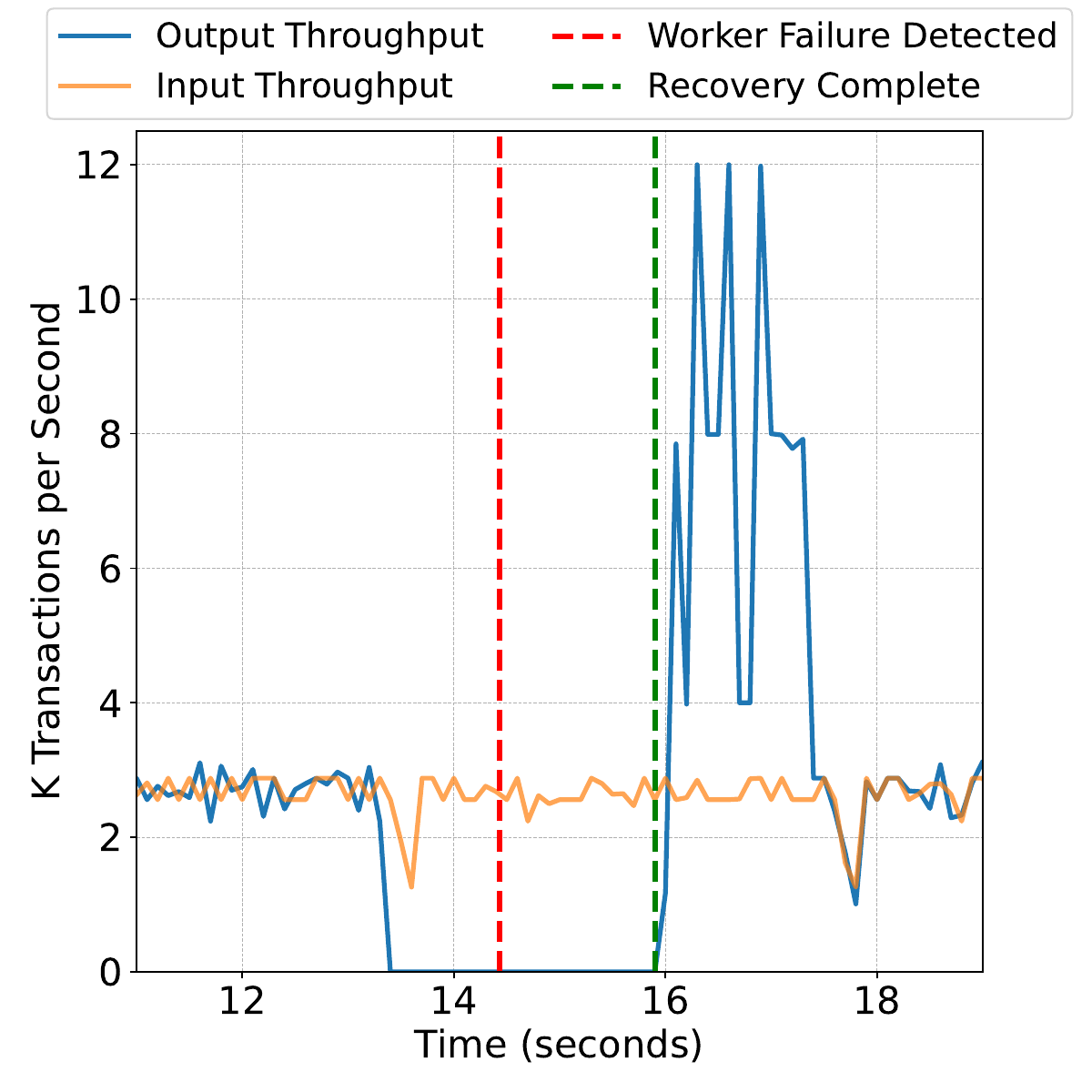}
            \caption{Throughput}
        \label{fig:exp-t-rec}
     \end{subfigure}
          \begin{subfigure}[b]{0.49\columnwidth}
        \centering
        \includegraphics[width=\columnwidth]{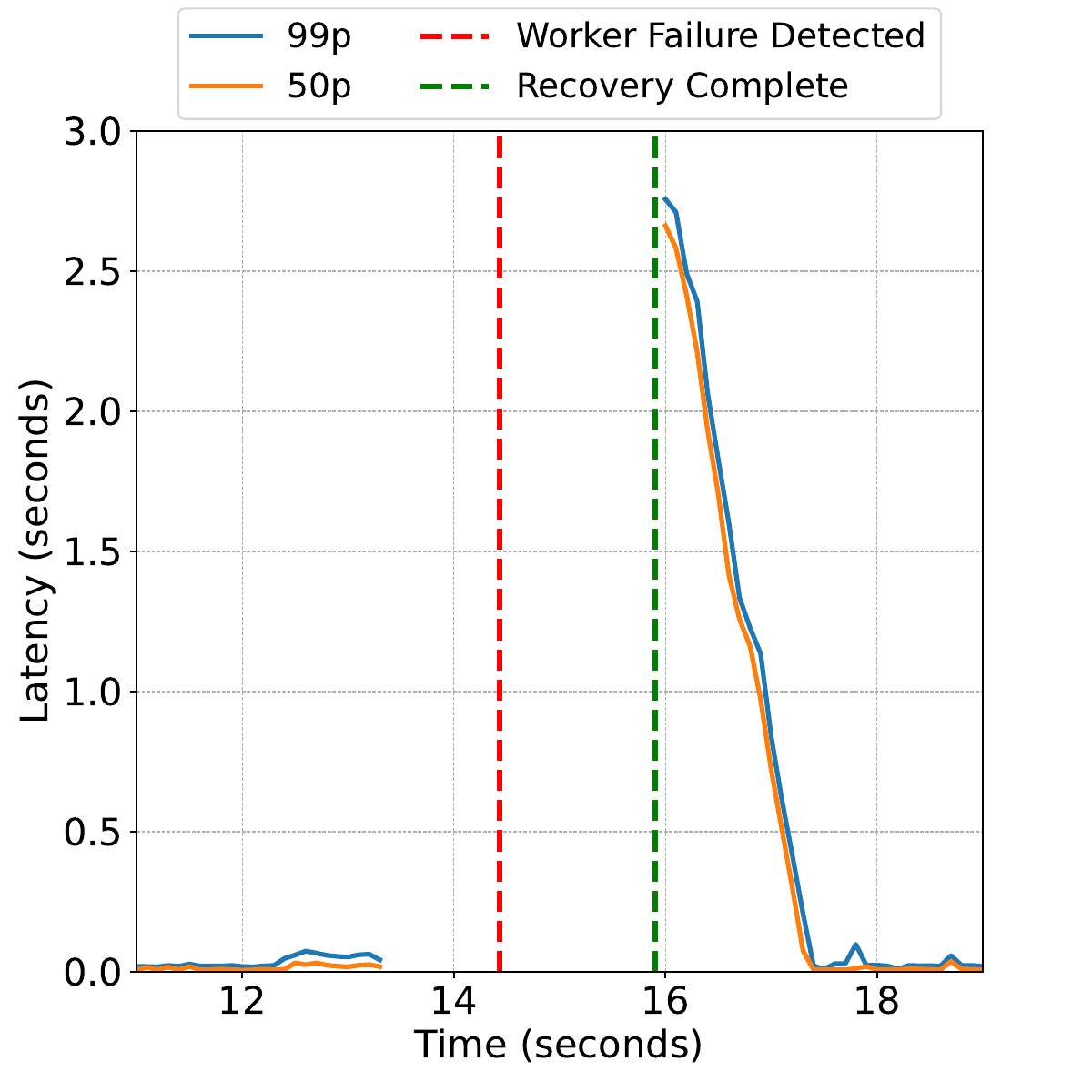}
            \caption{Latency}
        \label{fig:exp-l-rec}
     \end{subfigure}
        \caption{\sysname{}'s behavior during recovery.}
        \label{fig:exp-recovery}
\end{figure}


\para{Recovery Time} In \Cref{fig:exp-recovery}, we evaluate the recovery process of \sysname{} with the same parameters as in \Cref{fig:exp-snapshots}. We reboot a \sysname{} worker at \textasciitilde 13.5 seconds. It takes \sysname{}'s coordinator roughly a second to detect the failure. Then, after the reboot, the coordinator re-registers the worker and notifies all workers to load the last complete snapshot, merge any uncompacted deltas, and use the message broker offsets of that snapshot. The recovery time is also observed in the latency (\Cref{fig:exp-l-rec}) that is \textasciitilde 2.5 seconds (time to detect the failure in addition to the time to complete recovery). In terms of throughput (\Cref{fig:exp-t-rec}), we observe \sysname{} working on its maximum throughput after recovery completes to keep up with the backlog and the input throughput.

\begin{figure}[t]
    \centering
    \includegraphics[width=\columnwidth]{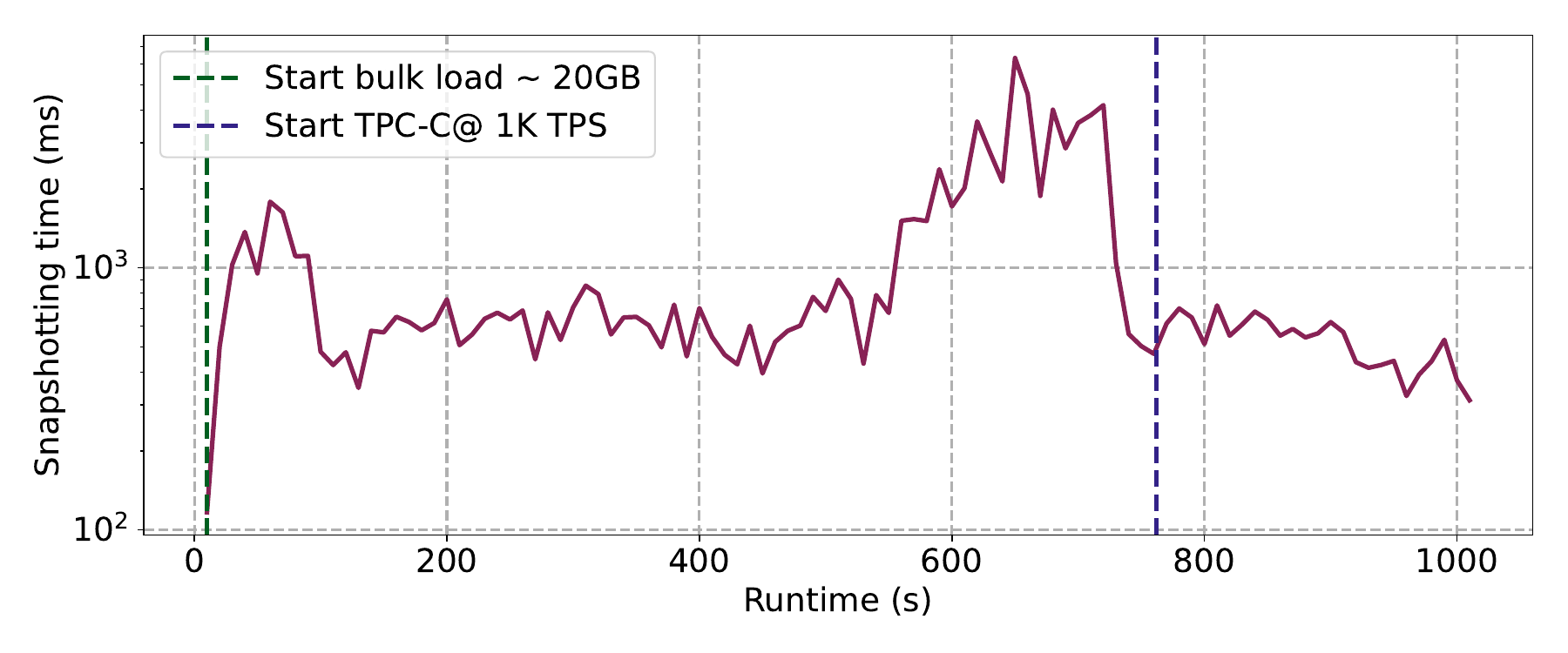}
    \vspace{-9mm}
    \caption{Behaviour of incremental snapshots on \sysname{} with \textasciitilde20GB TPC-C state.}
    \vspace{-4mm}
    \label{fig:incr_snap}
\end{figure}

\para{Effect of Large State Snapshots} In \Cref{fig:incr_snap}, we test the incremental snapshotting mechanism against a larger state of 20 GB from TPC-C using a bigger \sysname{} deployment of 100 1-CPU workers at 10-second checkpoint intervals. From 0 to the 750-second mark, \sysname{} is importing the dataset. Since there are no small deltas (importing is an append-only operation), snapshotting is more expensive than the normal workload execution, where only the deltas are stored in the snapshots. The increase in latency at \textasciitilde550 seconds corresponds to the loading of the largest tables (Stock and Order-Line) in the system. After loading the data and starting the transactional workload at 1000 TPS, we observe a drop in latency due to fewer state changes within the delta maps.

\section{Related Work} \label{sec:rel_work}
\para{Transactional SFaaS} SFaaS has received considerable research attention and open-source work. Transactional support with fault tolerance guarantees (that popularized DBMS systems) is necessary to widen the adoption of SFaaS. Existing systems fall into two categories: i) those that focus on transactional serializability and ii) those that provide eventual consistency. The first category includes Beldi~\cite{beldi}, Boki~\cite{boki}, and T-Statefun~\cite{tstatefun}. Beldi implements linked distributed atomic affinity logging on DynamoDB to guarantee serializable transactions among AWS Lambda functions with a variant of the two-phase commit protocol. Boki extends Beldi by adding transaction pipeline improvements regarding the locking mechanism and workflow re-execution. In turn, Halfmoon~\cite{qi2023halfmoon} extends Boki with an optimal logging implementation. T-Statefun~\cite{tstatefun} also uses two-phase commit with coordinator functions to support serializability on top of Apache Flink's Statefun. For eventually consistent transactions, T-Statefun 
implements the Sagas pattern. Cloudburst~\cite{cloudburst} also provides causal consistency guarantees within a DAG workflow. Proposed more recently, Netherite \cite{burckhardt2022netherite}  offers exactly-once guarantees and a high-level programming model for Microsoft's Durable Functions \cite{BurckhardtGJKMM21}, but it does not guarantee transactional serializability across functions. Unum \cite{LiuLNB23} needs to be paired with Beldi or Boki to ensure end-to-end exactly-once and transactional guarantees.


\para{Dataflow Systems} Support for fault-tolerant execution in the cloud with exactly-once guarantees~\cite{fernandez2014making, CarboneEF17} is one of the main drivers behind the wide adoption of modern dataflow systems. However, they lack a general and developer-friendly programming model with support for transactions and a natural way to program function-to-function calls. Closer to the spirit of \sysname{} are Ciel~\cite{murray2011ciel} and Noria~\cite{noria}. Ciel proposes a language and runtime for distributed fault-tolerant computations that can execute control flow. Noria solves the view maintenance problem via a dataflow architecture that can propagate updates to clients quickly, targeting web-based, read-heavy computations. However, none of the two provide a transactional model for workflows of functions like \sysname{}.

\para{Transactional Protocols} Besides Aria~\cite{aria} that inspired the protocol we created for \sysname{} \Cref{sec:dataflow-system}, two other protocols fit the requirement of no a priori read/write set knowledge: Starry~\cite{zhang2022starry} and Lotus~\cite{zhou2022lotus}. Starry targets replicated databases with a semi-leader protocol for multi-master transaction processing. At the same time, Lotus~\cite{zhou2022lotus} focuses on improving the performance of multi-partition workloads using a new methodology called run-to-completion-single-thread (RCST). \sysname{} makes orthogonal contributions to these works and could adopt multiple ideas from them in the future.




\vspace{3mm}
\section{Future Work}

\para{Elasticity in Dataflow Systems} Extensive work has been carried out in dynamic reconfiguration \cite{noria, ds2, dhalion} and state migration \cite{megaphone, meces, rhino} of streaming dataflow systems over the last few years. These advancements are necessary for providing serverless elasticity in the case of state and compute collocation to leverage dataflows as an execution model for serverless stateful cloud applications, which is a future goal of \sysname{}.

\para{Replication for High Availability} 
In the \sysname{} architecture, replication is only applied in the snapshot store and the Input/Output queues to ensure fault tolerance. For high-availability, \sysname{} could adopt replication mechanisms from deterministic databases. Specifically, the design of deterministic transaction protocols, such as Calvin \cite{thomson2012calvin}, feature state replicas that require no explicit synchronization. First, the sequencer replicas need to agree on the order of execution. After that, the deterministic sequencing algorithm guarantees that the resulting state will be the same across partition/worker replicas by all replicas executing state updates in the same order.

\para{Non-Deterministic Functions on Streaming Dataflows} In its current version, \sysname{} requires application logic to be deterministic, similar to OLTP~\cite{voltDB, one-shot-h-store}, where stored procedures are required to be deterministic since they run independently on different replicas. The same determinism requirement applies to SFaaS \cite{tstatefun, boki} systems. However, real-world applications may encapsulate logic that makes the outcome of their execution non-deterministic. Examples of non-deterministic operations are calls to external systems and using random number generators or time-related activities. \rev{That said, we have a plan for supporting non-deterministic functions in \sysname{}, as discussed in \Cref{sec:addr_non_det}}.

\section{Conclusion} \label{sec:conclusion}
This paper presented \sysname{}, a distributed streaming dataflow system that supports multi-partition transactions with serializable isolation guarantees through a high-level, standard Python programming model that obviates transaction failure management, such as retries and rollbacks. \sysname{} follows the deterministic database paradigm while implementing a streaming dataflow execution model with exactly-once processing guarantees. \sysname{} outperforms the state-of-the-art by at least one order of magnitude in all tested workloads regarding throughput.


\begin{acks}
We want to thank Paris Carbone for his advice throughout the process of developing \sysname{} and the anonymous reviewers for their constructive feedback.
This publication is part of project number 19708 of the Vidi research program, partly financed by the Dutch Research Council (NWO).
\end{acks}

\bibliographystyle{ACM-Reference-Format}
\bibliography{sample-base}

\end{document}